\documentclass[fleqn,usenatbib]{mnras}

\usepackage{newtxtext,newtxmath}
\usepackage{physics,amsmath}
\usepackage[normalem]{ulem}
\usepackage{xcolor}

\usepackage[T1]{fontenc}

\DeclareRobustCommand{\VAN}[3]{#2}
\let\VANthebibliography\thebibliography
\def\thebibliography{\DeclareRobustCommand{\VAN}[3]{##3}\VANthebibliography}

\usepackage{graphicx}	
\usepackage{amsmath}	

\usepackage{amssymb}	

\newcommand{\q}[1]{``#1''}

\newcommand{\rockstar}{\textsc{Rockstar}}
\newcommand{\ahf}{\textsc{AHF}}
\newcommand{\disperse}{\textsc{DisPerse}}

\newcommand{\gadgetx}{\textsc{Gadget-X}}
\newcommand{\hMpc}{{\ifmmode{\,h^{-1}\,{\rm Mpc}}\else{$h^{-1}$Mpc}\fi}}
\newcommand{\hMsun}{{\ifmmode{\,h^{-1}\,{\rm {M_{\odot}}}}\else{\,$h^{-1}\,{\rm{M_{\odot}}}$}\fi}}
\newcommand{\Msun}{{\ifmmode{\,{\rm {M_{\odot}}}}\else{$\,{\rm{M_{\odot}}}$}\fi}}
\newcommand{\hkms}{{\ifmmode{\,h^{-1}\,{\rm km\,s}^{-1}}\fi}}
\newcommand{\threehun}{{\sc The Three Hundred}}

\title[Filaments and shocks in galaxy clusters]{The three hundred project: thermodynamical properties, shocks and gas dynamics in simulated galaxy cluster filaments and their surroundings}

\author[Rost et al]{Agust\'{\i}n~M.~Rost$^{1, 2}$\thanks{E-mail: a\_rost@unc.edu.ar}, 
Sebasti\'an~E.~Nuza$^{3}$, 
Federico Stasyszyn$^{1}$, 
Ulrike Kuchner$^{2}$,
Matthias Hoeft$^{4}$,\newauthor
Charlotte Welker$^{7}$,
Frazer Pearce$^{2}$,  
Meghan Gray$^{2}$, 
Alexander Knebe$^{5, 6}$, 
Weiguang Cui$^{5, 6}$, 
Gustavo Yepes$^{5,6}$
\\
\\
$^1$ Instituto de Astronom\'ia Te\'orica y Experimental (IATE), Laprida 854, C\'ordoba, Argentina\\
$^2$ School of Physics \& Astronomy, University of Nottingham, Nottingham NG7 2RD, UK\\
$^3$ Instituto de Astronom\'{\i}a y F\'{\i}sica del Espacio (IAFE, CONICET-UBA), 1428 Buenos Aires, Argentina\\
$^4$ Th\"uringer Landessternwarte, Sternwarte 5, 07778 Tautenburg, Germany\\
$^5$ Departamento de F\'isica Te\'{o}rica, M\'{o}dulo 15, Facultad de Ciencias, Universidad Aut\'{o}noma de Madrid, 28049 Madrid, Spain\\
$^6$ Centro de Investigaci\'{o}n Avanzada en F\'{\i}sica Fundamental (CIAFF), Universidad Aut\'{o}noma de Madrid, 28049 Madrid, Spain \\
$^7$ NYC College of Technology, City University of New York, 300 Jay street, Brooklyn, NY, USA\\
}

\date{Accepted XXX. Received YYY; in original form ZZZ}

\pubyear{2022}

\begin{document}
\label{firstpage}
\pagerange{\pageref{firstpage}--\pageref{lastpage}}
\maketitle

\begin{abstract}
Using cosmological simulations of galaxy cluster regions from \threehun\ project we study the nature of gas in filaments feeding massive clusters. By stacking the diffuse material of filaments throughout the cluster sample, we measure average gas properties such as density, temperature, pressure, entropy and Mach number and construct one-dimensional profiles for a sample of larger, radially-oriented filaments to determine their characteristic features as cosmological objects. Despite the similarity in velocity space between the gas and dark matter accretion patterns onto filaments and their central clusters, we confirm some differences, especially concerning the more ordered radial velocity dispersion of dark matter around the cluster and the larger accretion velocity of gas relative to dark matter in filaments. We also study the distribution of shocked gas around filaments and galaxy clusters, showing that the surrounding shocks allow an efficient internal transport of material, suggesting a laminar infall. The stacked temperature profile of filaments is typically colder towards the spine, in line with the cosmological rarefaction of matter.  
Therefore, filaments are able to isolate their inner regions, maintaining lower gas temperatures and entropy. 
Finally, we study the evolution of the gas density-temperature phase diagram of our stacked filament, showing that filamentary gas does not behave fully adiabatically through time but it is subject to shocks during its evolution, establishing a characteristic $z=0$, entropy-enhanced distribution at intermediate distances from the spine of about $1-2\,h^{-1}\,$Mpc for a typical galaxy cluster in our sample.
\end{abstract}




\section{Introduction}
In the Universe there is an evident pattern of cluster regions containing hundreds of galaxies interconnected by filaments of several Mpc in length, which are, at the same time, bordering walls of galaxies of similar size. Furthermore, all these structures wrap around 
irregularly shaped cosmic voids \citep{DeLapparent1986, Bond1996, Colless2003}.

These large-scale structures result from the anisotropic gravitational collapse of random fluctuations in the beginning of the Universe, which grew since the Big Bang, and continue to do so in the present \citep[e.g.][]{Zeldovich1970,Peebles1980, Peacock1999, Planck2018}. This fact not only explains the spatial distribution of galaxies but also, to a great extent, the complex dynamics of the emerging cosmic web. 
In the latter, on multi-scale levels, galaxies generally flow from voids to flattened walls as a result of the gravitational collapse along one axis. They further collapse onto filaments which in turn transport galaxies into galaxy clusters that sit at the nodes of the largest filaments of the web-like large-scale structure. 
As a result, filaments are responsible for continuously feeding clusters with new galaxies \citep[e.g.][]{Zeldovich1970, vanHaarlem1993, Knebe2004, Cautun2014, Kuchner2022}, which end up orbiting these central objects eventually reaching a virial equilibrium or being destroyed by galaxy interactions. 
The dark matter and gas components follow a similar behaviour owing to the fact that they are influenced by the same gravitational field. Consequently, the gas permeating the Universe is also expected to flow from low to high-density regions, i.e. moving towards the potential wells determined by the evolving large-scale structures. However, while dark matter is collisionless and experiences only a gravitational pull and shell crossing, gas also exerts pressure and is able to cool down through radiative cooling and shocks, making the behaviour of these components diverge in high-density environments such as clusters of galaxies or inside haloes. These differences are evident when we consider, for instance, the formation of galaxy discs, stars, and shocked gas around haloes and galaxy clusters, among other phenomena that the dark matter does not exhibit.

As cosmological structure grows and evolves, strong shocks are expected to develop around filaments, where primordial gas is shock-heated for the first time \citep[][]{SunZel72, Miniati00}. Similarly, the collapse of gas surrounding galaxy cluster regions, where several filaments meet, can also give rise to the formation of shocks at large distances from the cluster centre. These so-called {\it accretion} shocks trace structure formation at large scales, typically reaching Mach numbers that can be well above 10 \citep[][]{ArayaMelo12, Power2020, Vurm2023} heating up the gas. Filaments have been proposed to host a significant fraction of the missing mass in the Universe in the form of a rarefied gas component \citep[][]{Cen99, Tuominen2021} known as the warm-hot intergalactic medium (WHIM). 
Its direct detection is challenging and has recently been achieved through measurements of individual galaxy cluster pairs \citep{Tittley2001, Werner2008, Akamatsu2017, Bonjean2018}, individual filaments \citep{Umehata2019, Bacon2021} or through the stacking of many filaments \citep{Tanimura2019, Vernstrom2021}. These accretion shocks have higher Mach numbers compared to those produced as a result of the hierarchical nature of halo growth, the so-called {\it merger} shocks \citep[][]{Hoeft11,Nuza12,Nuza17,Nuza23}, which provide additional gas heating to the intergalactic medium. 

Some works have studied the accretion of material onto the cluster by cosmic filaments \citep{Rost21, Kotecha2022, Vurm2023} and found that filaments continuously feed cluster centres with colder gas by resisting the pressure from the ICM. In particular, \cite{Rost21} (hereafter R21) compared gas accretion in clusters to that of dark matter. Although more dark matter is being accreted along the direction of filaments, its infall velocity seems not to depend on 
whether we look in the radial direction of the filament or not, in contrast to what happens with the gaseous counterpart. \cite{Kotecha2022} found evidence that filaments could {\it shield} galaxies to a certain degree from the more hostile intracluster medium as they enter the cluster environment well within the virial radius. These authors found the shielding effect to result from cooler, more collimated gas streams at the core of intra-cluster filaments. This is interesting because, although this work focuses on the internal cluster regions, for the outskirts the effect seems to be the opposite. There have been many studies finding that filaments have a significant role in the so-called ``preprocessing" of galaxies, affecting galaxy evolution and properties by diverse mechanisms such as ram-pressure, mergers and the like, inducing a quenching on the ability of galaxies to form stars before these objects enter the cluster environment \citep[e.g.][]{Wetzel2013, Martinez2016, Kraljic2018, Porter2008, Heines2015, Heines2017, Kotecha2022, Song2021}.

In this work, we take advantage of a sample of hydrodynamical galaxy cluster simulations performed within the context of \threehun\ project \citep{Cui2018}, a sample of 324 simulated galaxy cluster regions that allow us to study filaments at $z=0$ in cluster-like environments to characterise the kinematics of material in filaments, the distribution of shocks and their stacked density, pressure, temperature and entropy profiles. Moreover, we also study the mean phase-space temporal evolution of gas in the filament sample. To identify filaments in the simulations we use the DisPerSe filament finder including gas particles \citep{Sousbie2011}. This filament catalogue was previously used in R21, among others, and it allows us to infer statistical properties of filament gas dynamics in the neighbourhood of galaxy clusters. 

The paper is organized as follows. In Section~\ref{sec:simul} we describe the simulations analysed in this paper, the filament sample and the shock finder algorithm. In Section~\ref{sec:quantities} we describe the halo excision procedure and the filament stacking technique. In Section~\ref{sec:results} we present our results concerning mean gas properties and shock distribution in filaments at $z=0$, together with the stacked phase-space temporal evolution of filamentary gas. Finally, in Section~\ref{sec:discus}, we close the paper with a discussion and our main conclusions.

\section{Simulations}
\label{sec:simul}
\begin{figure*}
  \includegraphics[width=\textwidth]{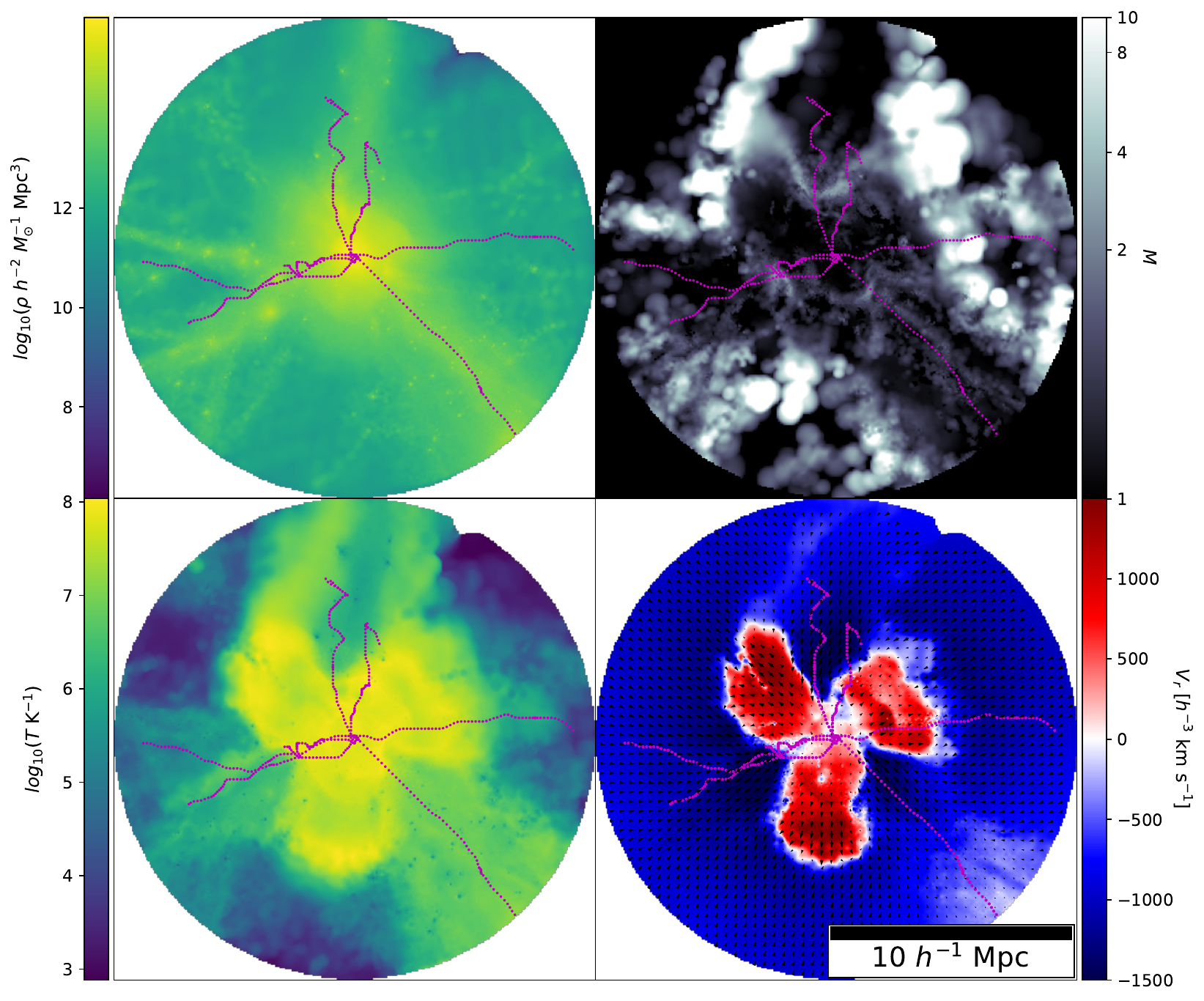}
  \caption{Gas distribution at $z=0$ around galaxy cluster $14$ centred in the simulated volume for a slice of $4\,h^{-1}\,$Mpc. Top left: logarithm of the mean baryon density in the pixel-column. Top right: mean mach number. Bottom left: logarithm of mean temperature. Bottom right: mean value of the radial component of the velocity with respect to the cluster centre without removing the infall velocity. Arrows represent the mean velocity in the plane of the plot and the purple dots represent the filamentary network identified by \disperse~(see text for details).
  }
  \label{fig:all1}
\end{figure*}

\subsection{The Three Hundred galaxy cluster sample}

\threehun\footnote{\url{https://the300-project.org}} \citep{Cui2018} simulations consist of $324$ spherical zoom-in re-simulations of galaxy clusters selected from the simulated MultiDark2 cosmological volume \citep{multidark} using the \rockstar\ halo finder \citep{rockstar}. The re-simulated regions were run with the code \gadgetx\ \citep{Springel2005, Murante2010, Beck2016, Rasia2015}, which implements full gas physics in a Lagrangian smoothed particle hydrodynamics (SPH) approach.
The particle resolution of the spherical volumes of radius $15 \hMpc$ consist of initial masses of $12.7 \times 10^8$ and $2.36 \times 10^8 \hMsun$ for the dark matter and gas components, respectively. These volumes aim to represent the $\sim 5R_{200}$\footnote{$R_{200}$ is the radius enclosing an overdensity of 200 times the critical density of the universe.} region around clusters at $z = 0$, as the outer layers of the spherical regions contain successively coarser SPH particles. The simulation runs are stored in $129$ snapshots and have been used to study the build-up of haloes in previous works \citep[e.g.][]{Arthur2019, Haggar2020}.

The haloes and subhaloes in the \threehun\ simulations were identified with the \ahf \footnote{\url{http://popia.ft.uam.es/AHF}} halo  finder \citep{AHF}, which determines several halo properties such as e.g. masses, angular momentum, luminosities, among other relevant quantities, taking into account all matter components, i.e. dark matter, gas and star particles. The range of cluster masses at the current time go from $M_{200} = 6.08 \times 10^{14}$ to $2.62 \times 10^{15} \hMsun$, and were selected as a mass-complete sample of the $324$ most massive virialised structures at $z=0$.

To illustrate our simulations, Fig.~\ref{fig:all1} shows a slice of $2\hMpc$ thickness for one of the re-simulations in the sample centred at the main galaxy cluster of the region. Different properties for the gas particles are depicted together with a network of filaments located in the simulated sphere as identified by the DisPerse code \citep{Sousbie2011}. The spatial distribution of these physical quantities were rendered with the software SPH-viewer \citep{sph-viewer}, which generates a two-dimensional (2D) array representing the distribution that takes into account the smoothing length of the particles. The averaged quantities were estimated by weighting the physical properties (e.g. temperature, radial velocity and Mach number) of the gas particles with their corresponding mass and then normalising them by the 2D array of the mass distribution, which was obtained using only the mass of the particles. For the density we determined the mass distribution by weighting by the particle's mass, and then dividing it by the corresponding volume of each cell. A more detailed description of the filament set in the simulated cluster regions will be given in Section~\ref{sec:fil_sample}. On the top left, the gas density is shown, naturally reaching its maximum at the galaxy cluster location, and also displaying clear filamentary structures consisting of three main branches converging at the centre. Additionally, it is also possible to spot less {\it contrasted} scattered within the simulated sphere that, given the {\it persistence} parameter used (a measure of the density contrast of the structures; see R21 for details), were not significant enough to not be discarded by the filament identification algorithm.
In the \threehun\ the persistence is chosen to focus on the most contrasted filaments, i.e. the ones that typically connect clusters of galaxies or large groups.

To estimate the distribution of shocked gas, we can look at the local Mach number of the gas particles. A more detailed description of the method used to determine this quantity in our simulations is outlined in Section~\ref{sec:shock_finder}.
On the top right panel, the distribution of Mach number $\mathcal{M}$ is shown for gas particles with $\mathcal{M} > 1$. This figure is produced assuming that, similarly to gas density, the Mach number values can be extrapolated using a smooth kernel around the particles. As seen in this figure, the shocked material is generally found wrapping the filaments around. Similarly, shocked gas is also located at a certain radius surrounding the central halo. However, in the inner cluster regions, merger and structure formation shocks outside filaments outskirts are also seen. 
The presence of shocks close to galaxy cluster outskirts is consistent with the findings of \citet{Vurm2023}. These authors found that filamentary gas flowing towards a simulated massive cluster would form a termination shock close to its virial radius, thus making these locations promising targets for the detection of WHIM gas in filaments with X-ray observations.

The temperature distribution is shown in the lower left panel. The underlying filamentary structures can roughly be spotted looking at the broad hot gas regions. Material at the boundary of the broad hot regions around filaments tends to have higher temperatures than in the spines and coincides with the location of the shocked gas envelope. Interestingly, where filaments meet the central cluster, dents in the roughly spherical distribution of hot material around the cluster systematically form, in a process analogous to the smaller-scaled ``cold streams'' studied by \citet{Dekel2009}.

Lastly, on the lower right panel, the velocity field is shown using arrows, with colour-coded regions according to the radial velocity component towards the centre. Blue colours represent infalling material, as expected for high-density regions, where gas is mainly channelled through the filaments towards the centre, while reddish colours correspond to outwardly expanding material. By visual inspection of the time evolution of this particular cluster region using the same techniques applied in \cite{Nuza17}, we have checked that the expanding gas seen in this panel is set in motion as a result of a recent multiple-merger event of the central cluster with several substructures, driving the formation of merger shocks in the inner cluster regions that happen to be spotted at the current time. Comparing the lower left and right panels, it can be seen that the outflowing gas is clearly correlated with the central hot gas clouds, sharing the same cavities seen in the temperature maps. Moreover, the hot regions seen at the edge of the red distribution in the velocity maps are the result of merger shocks raising the temperature behind the shocks as they mover through the ICM. An analysis of cluster merger shocks within the context of the \threehun\ project will be presented in a forthcoming paper (Nuza et al., in preparation). 

These accretion patterns, and other features, are consistent with the results of R21, where we studied the efficiency of gas transport towards clusters through filaments. In general, the trends discussed above can be found in other clusters of the simulation set, allowing us to perform, in the following sections, a statistical analysis of gas properties with focus on filaments and their adjacent regions.

\subsection{The filament sample}
\label{sec:fil_sample}

We use the same catalogue of filaments as in the work of R21. The latter consists of a set of 3D filaments for each one of the galaxy clusters in \threehun\ sample, obtained with the DisPerSe code \citep{Sousbie2011} from the distribution of gas particles. 

The DisPerSe software identifies many topological features of the cosmic web such as nodes, filament segments, saddle points, bifurcation points, among others. In particular, filaments correspond to ridges in the smoothed density field that connect local maxima (nodes) through a filament-type saddle point. Naturally, at these local maxima to which filaments are connected, dark matter haloes reside, which gravitationally dominate part of the filament up to the saddle point attracting material towards the nodes. As a result, these filaments are only half of the ``traditional'' ones studied in other works that, in fact, consider both their start and end as nodes. 
In order to extract DisPerSe filaments, the gas density distribution around each cluster was computed in a grid with cell size of $150\,h^{-1}\,$kpc in three dimensions. The filament search was performed within a cube of $30\,h^{-1}\,$Mpc on a side centred at the cluster's position. The grid was then smoothed out with a gaussian kernel of 8 cells, and it was used to extract the filament network with a persistence value of $0.2$. This value allows us to keep prominent filaments connecting clusters or groups of galaxies, and leave out thin tendrils.
In total there are $11058$ filaments distributed throughout the $324$ clusters. For a more detailed description of the filament identification procedure please refer to R21.

\subsection{The shock detection scheme}
\label{sec:shock_finder}

Shocks are identified in post-processing in the same way as in \cite{Nuza17} (see also \citealt{Nuza12} and references therein). For every gas particle, we define the {\it shock normal}, $\vectorbold*{n}\equiv-\nabla P/|\nabla P|$, by evaluating its pressure gradient and later impose a series of conditions to search for true shocks, i.e. the shocked region must be in a convergent flow and thermodynamic properties are required to increase when going from the upstream to the downstream region. Specifically, we demand, along the shock direction, that (i) the velocity field shows a negative divergence, (ii) the density increases from upstream to downstream and (iii) that the latter is also valid for the entropy.

To estimate the Mach numbers, we use the Rankine-Hugoniot jump conditions for hydrodynamical shocks \citep[see e.g.,][]{Landau59} for each one of the imposed conditions above and compute three Mach number values. For a conservative estimate of the Mach number we consider the Mach numbers derived in each shock and then take the minimum. For further details for our shock identification technique within SPH, we refer the reader to the papers mentioned above.

With this procedure we estimated shock Mach numbers for all gas particles in 279 out of the 324 galaxy clusters in \threehun\ sample, leaving out of the analysis only part of the less massive groups. This does not change the conclusions of our work as we have already large enough statistics. In what follows, all stackings for the gas properties of interest will be performed with clusters from this subsample.

\section{Analysis}
\label{sec:quantities}

The physical processes that occur in galaxy clusters, groups, filaments and other large cosmic structures determine the properties of baryons and galaxies inside them. One way to characterise their influence is by computing the distance to these structures, assuming that all objects under scrutiny have similar characteristics at a given position relative to them. Another general assumption is that haloes have influence on surrounding material up to a distance proportional to their $R_{200}$ radius. This, of course, depends on the scale considered, making the physical distance itself not a reliable parameter. In our sample, the influence of galaxy clusters goes up to $1-2\times R_{200}$ from the centre, where, for example, we can already see traces of the central halo's influence by looking at the proportion of backsplash galaxies \citep{Haggar2020}. Therefore, in what follows, we adopt distances normalised by $R_{200}$ as a better way to estimate any given halo's influence.

Following R21, we define, for each gas particle, the perpendicular distance to the closest filament, $d_{\rm fil}$, that is not closer to an endpoint. This is done in order to avoid nearby material being influenced by groups or clusters of galaxies residing at the termination of filaments. Once the closest filament is identified, we also define the distance along the spine starting from its node, $d_{\rm node}$ normalised by its $R_{200}$. This will likely be the central node for radial filaments, but it could also be a secondary halo in the simulation region. To better account for the central halo influence, we define $d_{\rm clus}$ as the distance to the centre of the cluster, for each particle, normalised by the central $R_{200}$.
Finally, we define the quantities $X$, $Y$ and $Z$ for a given particle and a roughly radial filament linked to the central node. These quantities are Cartesian coordinates in the frame of reference of the cluster, oriented along the considered filament, scaled
down by $R_{200}$ of the cluster. We determine these quantities so that the $X$ axis is aligned with the vector pointing from the node to the saddle point of the filament and, given an arbitrary rotation along this axis, we define the $Y$ and $Z$ axes to be perpendicular to each other. As a result, the $X, Y, Z$ coordinates correspond to the particle positions in the {\it filament-oriented frame}, so that the node and the saddle point of the filament would have $(0, 0, 0)$ and $(L/R_{200}, 0, 0)$ coordinates, respectively, where $L$ is the filament's length. It is worth noting that these coordinates do not follow the curvature of the filament axis like the quantities $d_{\rm fil}$ and $d_{\rm node}$.
To provide a hint about related physical quantities, the mean (median) $R_{200}$ of central clusters in our sample at $z=0$ is $1.56\hMpc$ ($1.523\hMpc$) with a mean (median) mass of $9.16\times10^{14}\hMsun$ ($8.22 \times 10^{14}\hMsun$).

\subsection{Halo excision}
\label{sec:haloexcision}

Cosmological filaments comprise a relatively smooth distribution of dark matter and gas that, at the same time, is populated by virialised regions where galaxies and groups are located. Inside these virialised structures small-scale physical processes take place that, although linked to the environment where haloes reside, do not necessarily characterise the environment itself as they are internal processes. For example, if we look at the gas content, one immediately notices the presence of a shock-heated gas corona surrounding haloes, in sharp contrast to the smoother, underlying WHIM of the filament.
As with R21, we consider these effects as a contamination from the small-scale processes present in haloes. So, in order to study the unpolluted diffuse gas from filaments, we proceeded to excise all particles inside spherical regions of $2R_{200}$ around haloes of masses above $10^{11}\,$M$_{\odot}$ (a discussion and justification for this choice is given in R21). Smaller haloes have intrinsic virial temperatures ($\log_{10}(T\,{\rm K}^{-1}) \approx 5.5 $) that are lower than the intrafilament gas one, thus not affecting the overall WHIM distribution.
After the excision process, about $70\%$ of the gas is removed from the sample. 

\begin{table}
\caption{Properties of the filament samples discussed in Section~\ref{sec:parallel-perpendicular}. From top to bottom the rows show the number of filaments; their mean length in $h^{-1}\,$Mpc, and the mean angle of filaments in degrees with respect to the radial vector from the cluster centre.}
\begin{center}
\begin{tabular}{||c c c c c||} 
 \hline
 Sample: & All & Parallel & Perpendicular & Prominent  \\ [0.5ex] 
 \hline\hline
 $N$ & 9638 & 1242 & 858 & 574 \\ 
 \hline
 $\langle L \rangle$ & 3.77 & 3.14 & 2.67 & 9.04 \\
 \hline
 $\langle \alpha_{\rm fil} \rangle$ & 39.33 & 16.56 & 76.33 & 27.28 \\  
 \hline
 \label{tab:properties}
\end{tabular}
\end{center}
\end{table}

\subsection{Stacking procedure}
\label{sec:stacking}

To statistically study the effects observed in Fig.~\ref{fig:all1} we stack the data of our cluster sample using the quantities defined in Section~\ref{sec:quantities}. 
The value of each bin is determined using a 2D histogram weighted by gas particle mass times the quantity of interest (i.e. temperature, entropy, radial velocity and Mach number). For example, in the case of temperature, the following equation applies

\begin{equation}
I_{T}(i, j) = \log_{10}\left(\frac{\sum_{k \epsilon {\rm bin}} T_{k} m_{k} }{\sum_{k \epsilon {\rm bin}} m_{k}}\right){\rm ,}
\end{equation}

\noindent where $I_{T}(i, j)$ is the temperature in bin $(i, j)$ which is constructed using spatial coordinates, such as e.g. $d_{\rm fil}$, $d_{\rm clus}$, $x$, $y$, and the sum is performed for all particles falling in the current bin for the entire cluster sample. To determine the density of each bin, their volume was estimated using a distribution of random points imposing a mask that excludes regions around haloes to mimic the distribution of considered particles as explained in the previous section. Note that it is possible that not all filaments will contribute to the signal in the same way for all $d_{\rm fil}$ and/or $d_{\rm clus}$ explored values. However, when stacking a sample of multiple filaments, empty bins will fill up, although not evenly across the full spatial range. To take this into account, in order to assess the reliability of the signal, we also indicate the bin counts of each stacking using contours lines in some plots.

\subsubsection{Shocks in `parallel' and `perpendicular' filaments}
\label{sec:parallel-perpendicular}

By inspecting several figures like Fig.~\ref{fig:all1} for the rest of the clusters, we observe that the Mach number-enhanced regions around filaments is a common feature of these structures regardless of their orientation.
To check for this, we split the filament sample in those oriented radially towards the centre of the cluster ({\it parallel} filaments), and secondary filaments which are rather perpendicular to the radial direction ({\it perpendicular} filaments).

\begin{figure}
  \includegraphics[width=\columnwidth]{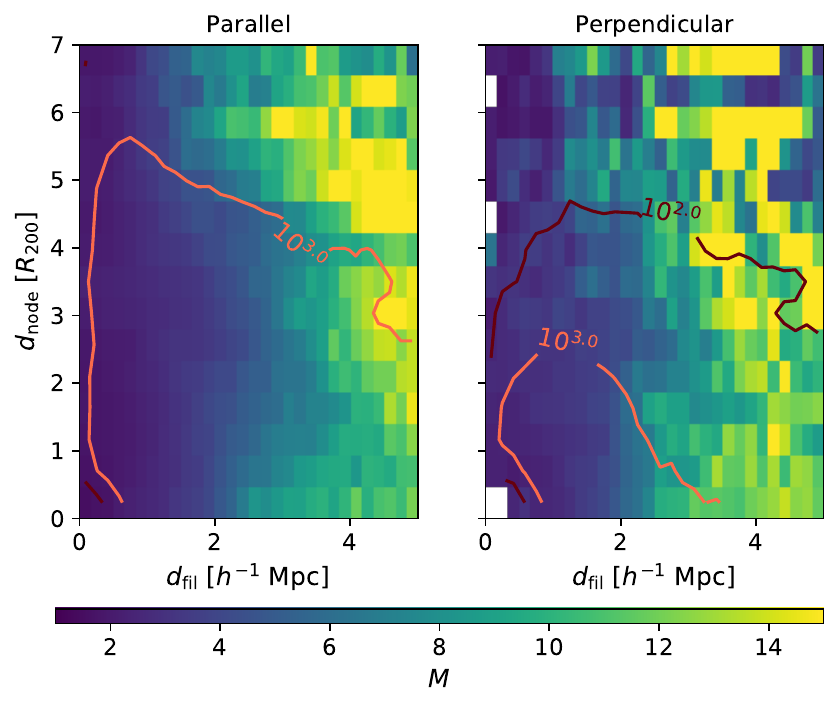}
  \caption{Stacking of the Mach number for shocked gas in filaments. Particles whose closest filaments is radially aligned with the radial direction from the centre of the cluster are shown on the left (see the explanation of filament classification in the text), whereas those that are perpendicular are shown in the right. The tolerance angle in each case is set to $20^{\circ}$. Contours indicate particle counts per bin.}
  \label{fig:parallel-perpendicular}
\end{figure}

To properly define these samples, we compute, for each filament, the cosine of the angle of its segments and the radial direction (see below) and average them to assign an angle $\alpha_{\rm fil}$ to the filament as follows:
\begin{equation}
    \cos\alpha_i = \frac{({\bf r}_{n + 1} - {\bf r}_{n}) \cdot \hat{\bf{ u}}_{n} }{|{\bf r}_{n + 1} - {\bf r}_{n}|},
\end{equation}
\begin{equation}  
    \langle\cos\alpha_{\rm fil}\rangle = \frac{1}{L}\sum_{n = 1}^{N - 1} \cos\alpha_{i}|{\bf r}_{n + 1} - {\bf r}_{n}|,
\end{equation}
where ${\bf r}_{n}$ is the position to the $n$th node, $\hat{\bf{u}}_{n}$ is the direction from the centre of the cluster to the centre of the segment, $\alpha_{i}$ is the angle with the radial direction of the $i$th segment, $|{\bf r}_{n + 1} - {\bf r}_{n}|$ is the length of the segment, which all add up to $L$, the total length of the filament.
With this method we obtain, for each filament, an angle $\alpha_{\rm fil}$ as an average of the perpendicularity of the filament's segments with respect to the radial vector from the cluster centre and use this angle to classify the filament sample. In particular, those filaments with $\alpha_{\rm fil} \leq 20^{\circ}$ are classified as parallel ($1242$ out of $9638$ filaments in the sample), whereas those satisfying that $\alpha_{\rm fil} \geq 70^{\circ}$ are perpendicular ($858$ in total). 
Filaments that are too bent or warped would unlikely be classified as parallel regardless of their orientation. However, we consider this rough approximation of perpendicularity good enough for our purposes.
Some of the properties of our filament samples are listed in Table~\ref{tab:properties}. Apart from the intrinsic difference in filament orientation, the two samples differ in filament length, with perpendicular filaments being shorter on average. The column referring to `All' considers the 9638 filaments in our sample of cluster regions with Mach number estimates and then comprise a subset of the sample of 11058 \disperse\ filaments identified in the total sample of 324 simulated clusters. The remaining 1420 filaments were thus not considered in this work.

\begin{figure}
  \includegraphics[width=\columnwidth]{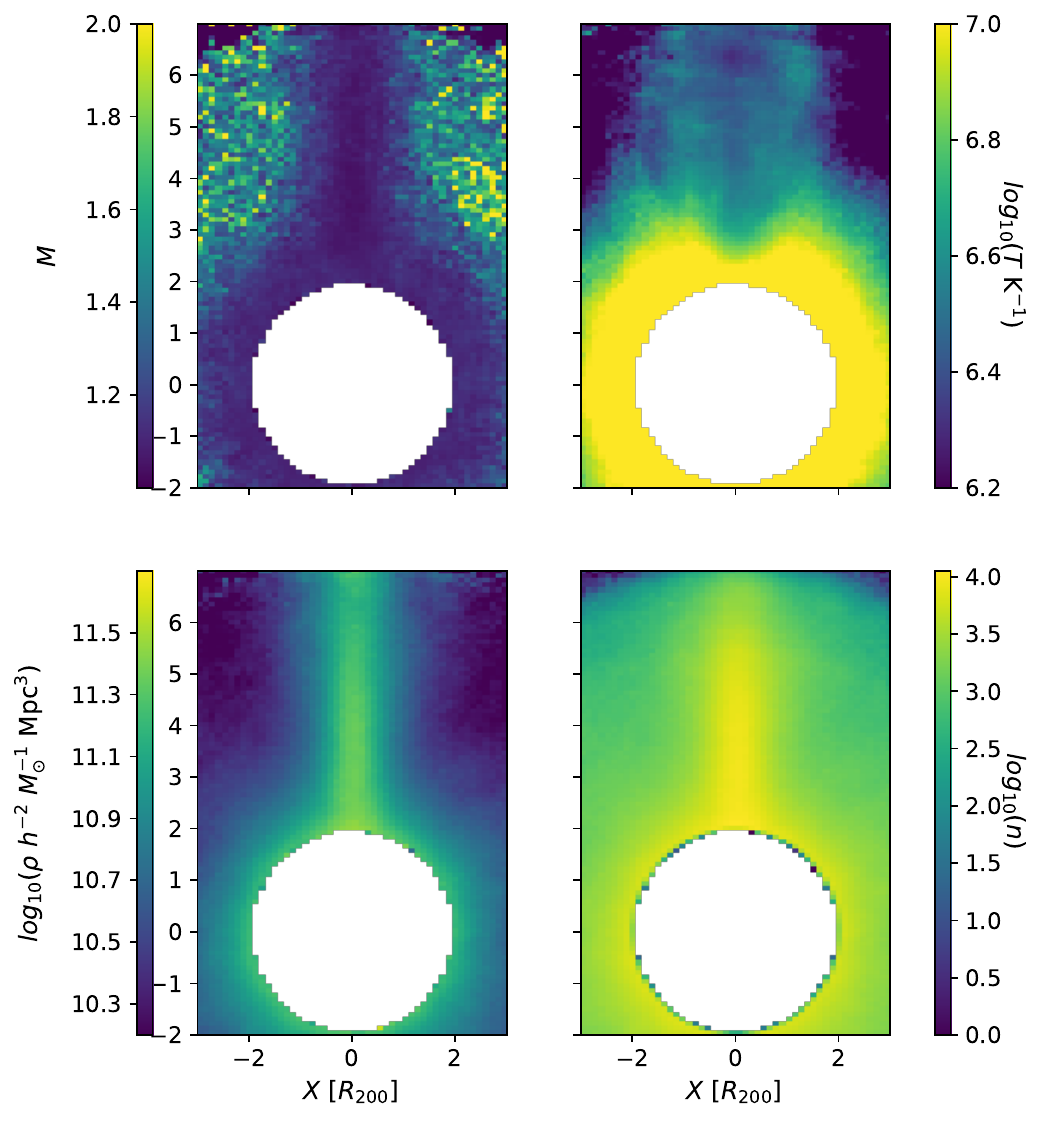}
  \caption{Stacked sample for all filaments at $z=0$ connected to the central cluster that are longer than $5\,h^{-1}\,$Mpc. Top left: mean Mach number. Top right: gas temperature. Bottom left: gas density. Bottom right: particle count. All gas particles inside $2 R_{200}$ were removed.
}
  \label{fig:alternative-stack}
\end{figure}

In Fig.~\ref{fig:parallel-perpendicular}, we stack the Mach number of gas in our filament sample in the $d_{\rm clus}-d_{\rm fil}$ plane following the procedure outlined in Section~\ref{sec:stacking}, with the parallel filaments on the left-hand panel, and the perpendicular on the right. The number of particles in each bin is shown using contour levels. The perpendicular sample is smaller than the parallel one and their filaments are {\it weaker}, which explains the order of magnitude difference in the amount of stacked particles seen in the plot, and the consequently noisier signal of the perpendicular filaments.
However, for both cases, the Mach number tends to be lower near the stacked filament core.
The widening of filaments at shorter distances from cluster centre is evident in the parallel sample. 
Since these filaments are approximately radial, the $d_{\rm clus}$ distance can be roughly interpreted as filament elongation. This feature is also seen in the perpendicular sample although, in this case, the interpretation is more tricky since $d_{ \rm clus}$ roughly corresponds to the radial location of the whole filament. In both cases, however, there appear shocks surrounding the central spines and we do not see much difference between the two samples, despite of the noisier distribution of the perpendicular sample owing to its smaller number size ($858$ vs. $1242$ filaments).   

\subsubsection{The sample of `prominent' filaments}

Although the presence of shocked gas surrounding filamentary structures does not seem to be unique to the radial filaments, the perpendicular sample is smaller and contains less significant filaments. Furthermore, their position with respect to the central cluster will vary in the $d_{\rm fil}$ vs. $d_{\rm node}$ diagram, adding an additional difficulty when stacking these structures. For these reasons, from now on, we will focus only on a sample of ``prominent'' radial filaments defined by all structures starting within the $R_{200}$ of the central cluster extending to at least $5\,h^{-1}\,{\rm Mpc}$ away, which for our cluster sample corresponds to lengths within the range $2.2-3.57 ~ R_{200}$. In this case, the resulting filaments are significantly longer than those considered in the parallel and perpendicular samples. Furthermore, filaments in the prominent sample are rather aligned with the radial orientation, as it can be seen when comparing their average angle with that of the total sample. The properties of the prominent sample are summarised in Table~\ref{tab:properties}.

\section{Results}
\label{sec:results}

\subsection{The stacked filament}
\label{sec:stackedfilament}

In Fig.~\ref{fig:alternative-stack}, we plot all prominent filaments as a unique stacked structure in the vertical direction by conveniently rotating the clusters, using the coordinates $(X, Y)$ introduced in Section~\ref{sec:quantities}, as explained in Section~\ref{sec:stacking}. For this {\it stacked filament} we studied the Mach number, temperature, density and particle counts by removing all particles inside $2 R_{200}$ to focus on filament properties only. All in all, after selecting the prominent parallel structures in the sample, a total of $574$ filaments have been stacked out of a total of $9638$ filaments found in our filament sample (see Section~\ref{sec:parallel-perpendicular}). The last quantity provides an impression of the available sampling in the stacked cluster regions.  

The stacked Mach number distribution is shown in the upper left panel. Two salient features can be immediately noticed. First, the central cluster region appears to be surrounded by shocked gas that most likely comprise the known accretion shocks onto clusters. Similarly, an increase of the Mach number with the distance to the filament is seen in a region that wraps the structure (similar to Fig.~\ref{fig:parallel-perpendicular}), close to the temperature bump regions. These shocks could coincide with the location of the edges of the rich vorticity regions of filaments, studied by \citet{Song2021}. Second, the central filament spine and its surroundings is essentially devoid of shocks directly feeding the central cluster with fresh WHIM gas as a consequence.   

As for the temperature (see right upper panel) there is a clear hotter region corresponding to the central cluster temperature, whereas the stacked filament shows two mild temperature bumps when traversing the filament along the horizontal direction surrounding a colder spine.
These bumps in temperature occur at the edges of the hot gas associated to the filament. Similar features were also reported for smaller-scale filaments in the works of \citet{Klar-Mucket2012} and \citet{Ramsoy2021}, where the temperature increase observed is the result of the formation of shocks around the filaments. Interestingly, the filament spine reaches further towards the centre in comparison with any other direction from the origin suggesting that cooler material travelling through the filament can, in fact, carve the ICM in its inward journey. This effect can also be seen in the distribution of the colder gas as a widening of the region when we travel away from the cluster.  
The density structure of the resulting stacked filament can be seen in the lower left panel having, as expected, a well-defined higher-density spine that vanishes with the distance from the axis. The density distribution surrounding the central cluster region is also observed as a an increase of the density as we approach the excised volume.

Finally, we can asses the reliability of the signal by inspecting the particle count in the bottom-right panel of Fig.~\ref{fig:alternative-stack}. The majority of the bins have accumulated above $10^2$ particles, which we consider high enough to not be considered noise. Note that this quantity is related but not equivalent to the density, to calculate the last, we require the estimation of the volume by means of randomly distributed particles.

\subsubsection{Filament profiles}

For our quantitative analysis of the filament sample studied in the previous subsection at $z=0$, we first estimate the average filament density profiles by fitting them with a mathematical model.
Owing to the low dependency of the profile on the $Y$-axis 
position, we integrate the density profile (lower left panel in Fig.~\ref{fig:alternative-stack}) along the $X$-axis in the range $[2.5, 7]$ in $R_{200}$ units and, taking advantage of the mirror symmetry, we stack both sides and fit the gas density profile with the following function:
\begin{equation}
    \rho(r) = \frac{\rho_{0}}{1 + (\frac{r}{r_{0}})^{\beta} } + \rho_{b}.
    \label{Gaus_rho}
\end{equation}
This profile has been used in other works like e.g., \citet{Colberg2005} and \citet{Galarraga-Espinosa2020}. To find optimal parameter values we use a nonlinear least squares method. For the gas density we found $\rho_{0}=1.76\times10^{11}\,h^{2}\,$M$_{\odot}\,$Mpc$^{-3}$, $\rho_{\rm b}=1.13\times10^{10}\,h^{2}\,$M$_{\odot}\,$Mpc$^{-3}$, $\beta = 2.12$ and $r_{0} = 0.4\,R_{200}$. In this case, there is also a mild dependence along the filament that we take into account as an uncertainty for the fitting. 

\begin{figure}
  \includegraphics[width=\columnwidth]{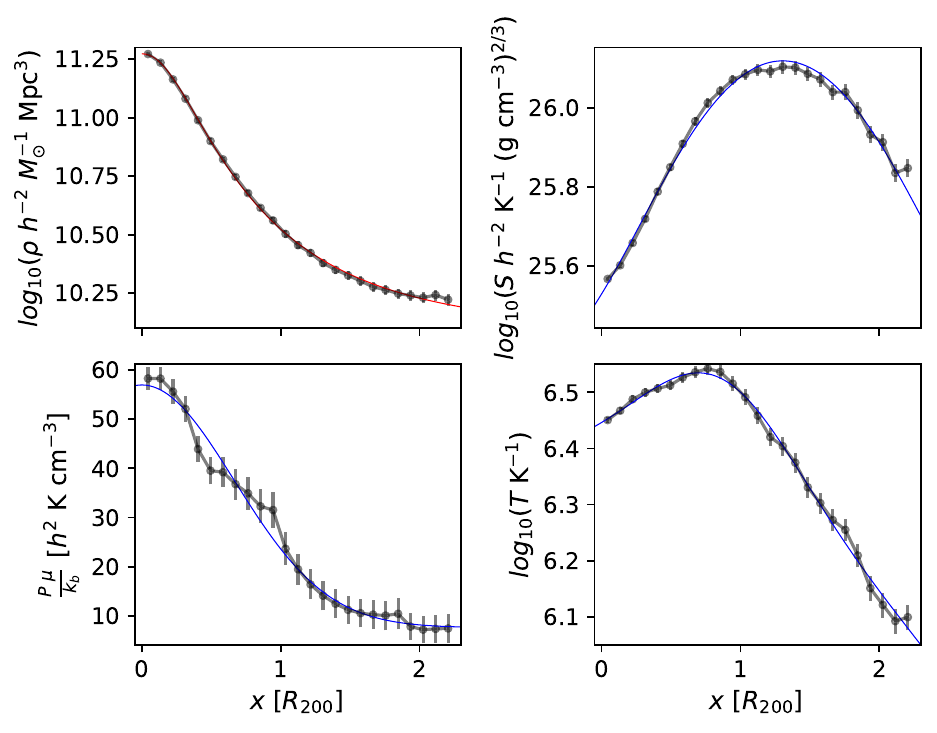}
  \caption{Radial profiles for gas density, entropy, pressure and temperature of the stacked filaments shown in Fig.~\ref{fig:alternative-stack} (solid circles) and their corresponding best fitting functions (see text).}
  \label{fig:fit}
\end{figure}

We also define an ``entropic function'' as $S = T\rho^{-\frac{2}{3}}$ (not shown in Fig.~\ref{fig:alternative-stack}), which we stacked in the same way as the other physical quantitites to obtain a 1D profile. In this case, we adopt a Gaussian function with an offset for the fit, namely
\begin{equation}
    S(r) = A_{\rm b} + A_{\rm e} ~ e^{-\frac{1}{2} \left(\frac{r-r_0}{\sigma}\right)^2},
    \label{Gaus_entropy}
\end{equation}
\noindent with best-fitting values $A_{\rm b} = 1.97\times10^{25}$, $A_{\rm e} = 1.12\times10^{26}$ (both in units of $h^{2}\,{\rm K}\,({\rm g}\,{\rm cm}^{-3})^{-2/3}$), $\sigma = 0.64\,R_{200}$ and $r_0 = 1.30\,R_{200}$. Similarly to the entropic function, we also estimated the pressure per particle as $P\mu/k_{\rm b}=\rho T/m_{\rm p}$ in units of $h^{2}\,{\rm K}\,{\rm cm}^{-3}$, where $m_{\rm p}$ is the proton mass and $\mu$ is the mean molecular weight in units of $m_{\rm p}$. To fit this quantity, we use a similar formula as that of Eq.~(\ref{Gaus_entropy}) centred in the filament spines (i.e., we take $r_{0} = 0$), obtaining best-fit values of $A_{\rm b} = 7.62$, $A_{\rm e} = 49.32$ (both in units of $h^2\,{\rm K}\,{\rm cm}^{-3}$) and $\sigma = 0.67\,R_{200}$.

For the gas temperature profile, a more complex function is needed, that we take by adding two Gaussian profiles. 
This function allows us to capture the decrease in temperature towards the spine, as well as a characteristic radii from which the profile starts to decrease as one moves away from the filament, i.e.
\begin{equation}
    \log_{10}(T(r)\,{\rm K}^{-1}) = A_{1} e^{-\frac{1}{2} \left(\frac{r}{\sigma_1}\right)^2} + A_{2} e^{-\frac{1}{2}\left(\frac{r - r_0}{\sigma_2}\right)^2}.
    \label{Gauss_temp}
\end{equation}
The best-fitting values found in this case are $A_{1} = 6.40$, $\sigma_1 = 6.83\,R_{200}$, $A_{2} = 0.18$, $r_0 = 0.85\,R_{200}$ and $\sigma_2 = 0.53 \,R_{200}$.

Fig.~\ref{fig:fit} shows the stacked profiles for the gas properties discussed above and the resulting fits. The profile scale-lengths can be accordingly computed for any cluster. For instance, using the average $R_{200}$ of the cluster sample given in Section~\ref{sec:quantities} ($\approx1.5\hMpc$), it is possible to estimate the typical locations of temperature, pressure and entropy bumps at about $1-2\hMpc$ from the filament spines. 
If we assume a typical value of $R_{200}\approx1.5 \hMpc$ for the gas density profile ($r_0=0.4~R_{200}$), and compare our results with the work of \citet{Galarraga-Espinosa}, we found $r_{0} \approx0.88\,$Mpc (using $h=0.678$), which is higher than the values $r_0=0.51-0.56\, {\rm Mpc}$ found by these authors for filaments selected from the IllustrisTNG simulation \citep{Nelson2019}. However, in our work, we are mainly focused in filaments located in high-density regions that are connected to quite massive galaxy clusters with $M_{200}\sim10^{15}\,$M$_{\odot}$, which may explain the differences. For the exponent of the gas density profile, \citet{Galarraga-Espinosa} obtained a value of $\alpha\approx 1.53-1.57$ for their filament sample (see their Eq.~(5)) whereas, in our case, we found $\beta = 2.12$.

In relation to the gas pressure profiles, our results are in line with those found by \citet{galarraga-espinosa2021} in the IllustrisTNG cosmological box for their `short' filaments which are more comparable in size to our prominent filament sample, although our values are somewhat higher. This could also be explained by the intrinsic higher environmental temperatures we are sampling, and/or by the implementation of the different feedback models.

Finally, we note that an unbiased comparison between our results and other simulations is not possible as the former is affected by the different numerical techniques, average procedures of the filament samples and the maximum radial distances considered. In our case we explored distances up to about $\gtrsim3\hMpc$ from the filament spines, whereas \citet{Galarraga-Espinosa} perform their sampling up to about $100\,$Mpc.

\subsection{Angular dependence}
\label{sec:angular}

\subsubsection{Filament gas properties}
\label{sec:filamentgasprop}

To take into account the approximately spherical symmetry of the resimulated galaxy clusters, we also perform the stacking of prominent parallel filaments by measuring the angle between the direction of a filament seen from the centre of the cluster and the direction to the gas particles.

\begin{figure}
  \includegraphics[width=\columnwidth]{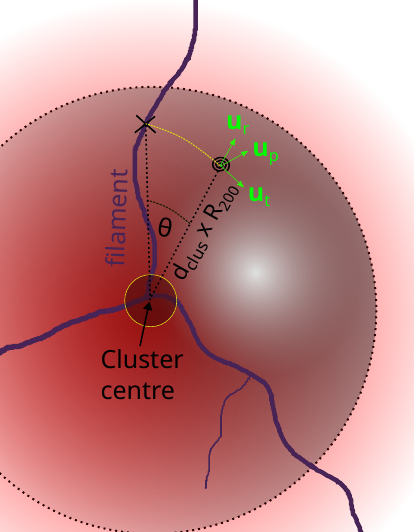}
  \caption{Diagram of the angular quantities defined in Section~\ref{sec:filamentgasprop}. The spiral represents a particle or halo to be studied, which is embedded in a sphere whose radius is the particle's distance to the centre of the cluster (translucent sphere). The filament network is represented in purple, and the unit vectors used to define velocities (see section \ref{sec:vel_fields}) are shown in green. The filament crosses the sphere at the location marked with \q{X}, which is linked to the location of an arbitrary particle by the yellow arc. The angle $\theta$ is then defined as the angle subtended by the yellow arc seen from the cluster centre.}
  \label{fig:diagram}
\end{figure}

For this statistic, we determine all filaments intersecting a spherical surface of radius $d_{\rm clus}$, measure the angles $\theta_i$ between the intersecting points ${\bf r}_{\rm x}$, the centre of the cluster ${\bf r}_{0}$, and a given gas particle located on the surface of such sphere ${\bf r}_{\rm p}$ so that

\begin{equation}
    \cos(\theta) = \frac{({\bf r}_{\rm x} - {\bf r}_{0}) \cdot ({\bf r}_{\rm p} - {\bf r}_{0})}{|{\bf r}_{\rm x} - {\bf r}_{0}||{\bf r}_{\rm p} - {\bf r}_{0}|},
\end{equation}

\noindent and finally select the filament for which $\theta_i$ is minimum. Fig.~\ref{fig:diagram} shows an illustration of this procedure. In this way, we can isolate the most likely filament linked to the gas particle of interest. Using this measure, the vertical line $\theta=0$ represents the stacked filament spine from $d_{\rm clus}=0$ to $7R_{200}$. Similarly, material that falls close to $\theta = 90^{\circ}$ will be located perpendicular to it. Note that, in this case, not all filaments would contribute to this signal at these distances since it is possible for a cluster to have a significant amount of them emanating from the centre. In such a case, when trying to leave a filament's influence zone, one instead falls under the influence of another neighbouring filament (at a $\theta$ ranging from $\approx 50 ^\circ$ to $\approx 90 ^\circ$, assuming connectivities in the range $8-20$, \citet{Codis2018}), thus the former will only contribute to the signal up to a certain angle only. 

\begin{figure}
  \includegraphics[width=\columnwidth]{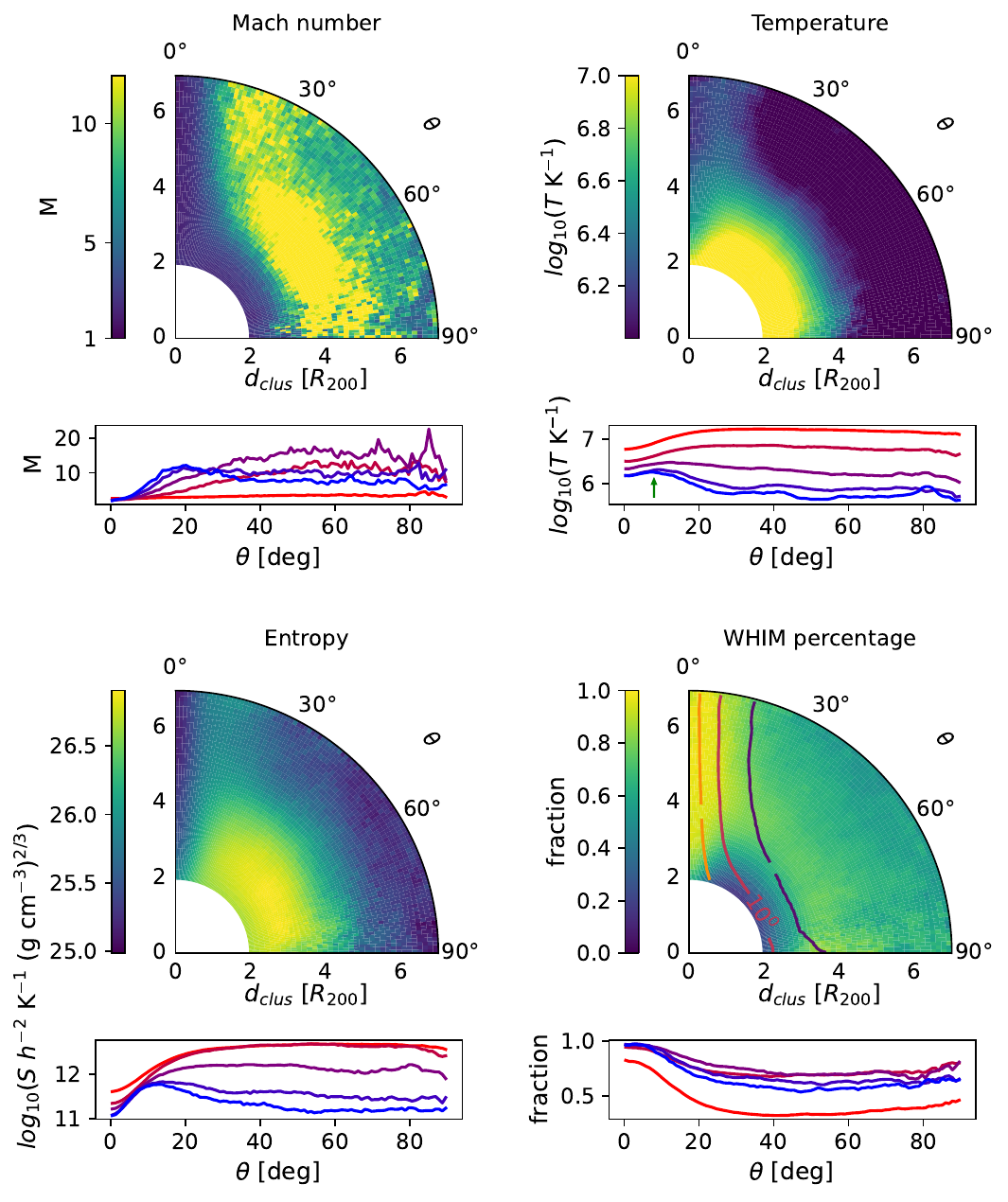}
  \caption[ArcDistanceNonShocked]{
  Arc-distance stacking of all gas particles around filaments connected to galaxy clusters at distances $[2,7]\times R_{200}$. From top left to bottom right: Mach number, temperature, entropy and WHIM fraction. In the lower right, contour lines indicate gas density.
  The profiles below each panel show the plotted quantities along the separation angle $\theta$ at different radii from $d_{\rm clus} = 2R_{200}$ (red lines) to $7R_{200}$ (blue lines). The green arrow in the upper-right panel indicates the approximate location of the temperature bump.
}
  \label{fig:Stack_all}
\end{figure}

\begin{figure}
  \includegraphics[width=\columnwidth]{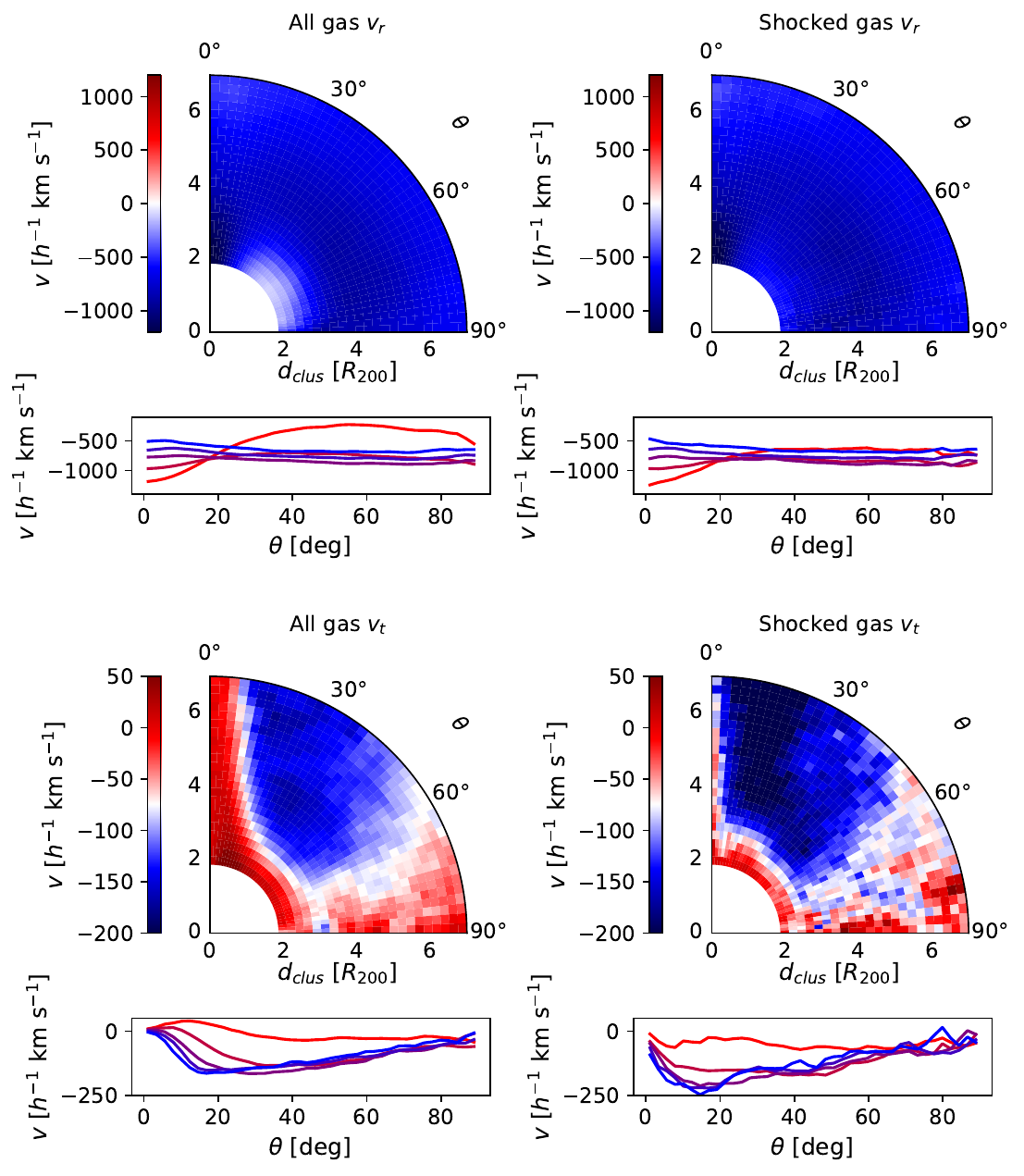}
  \caption[ArcDistanceShocked]{
   Arc-distance stacking of the radial (upper row) and tangential (lower row) velocity components defined in Section~\ref{sec:vel_fields} for all gas particles around filaments connected to galaxy clusters at distances $[2,7]\times R_{200}$. Quantities corresponding to all (shocked) gas is shown in the left (right) column. 
   The profiles below each panel show the plotted quantities along the separation angle $\theta$ at different radii from $d_{\rm clus} = 2R_{200}$ (red lines) to $7R_{200}$ (blue lines).}
  \label{fig:Stack_shocked}
\end{figure}

In Fig.~\ref{fig:Stack_all} we show different stacked physical quantities as a function of $d_{\rm clus}$ and $\theta$ for all particles that fulfil the conditions listed above, whereas Fig.~\ref{fig:Stack_shocked}, which displays gas velocity components, only considers those that are currently being shocked (i.e. $\mathcal{M} > 1$). Additionally, all panels include the one-dimensional (1D) tangential profiles at different radii as a function of $\theta$. The curves are colour-coded according to radial distance with the red (blue) end representing the inner (outer) profiles from $2$ to $3$ ($6$ to $7$) in units of $R_{200}$.

The upper left panel of Fig. \ref{fig:Stack_all} shows the stack of the mean Mach number. As noticed in the previous plots, much lower Mach numbers are seen inside the filaments compared to the surrounding ``shocked gas envelope'' at $10^{\circ} \lesssim \theta \lesssim 30^{\circ}$ and non-filament regions. This is clearly seen in the 1D profiles for all radial distances. In general, Mach numbers tend to decrease at $d_{\rm clus}\lesssim 3~R_{200}$, whereas between $3~R_{200}$ and $4~R_{200}$ a large shocked region centred at approximately $60^{\circ}$ is seen. Additionally, the 1D profiles show a shift of the Mach number peak towards the filament as we get further away from the cluster.

The stacked temperature distribution is shown in the right upper panel displaying a radial increase of the temperature towards the cluster centre, where most of the hotter material is located. Naturally, hot gas associated to the filaments extending for couple of Mpc from the filament axis is also seen. This is observed in many individual cases as a weak increase of temperature at the edge of a filament-like structure, either detected or not, by the filament finder. This seems to occur to some filaments at different distances from the axis. Hence, the trend is a bit blurred out in the stacking. As pointed out in the previous section, a lower-temperature cavity carving its way along the approximate direction of the filament spine near the cluster is evident at $2 R_{200} \lesssim d_{\rm clus} \lesssim 3R_{200}$, which was also observed by \citet{Klar-Mucket2012} in an idealised gas filament simulation at smaller scales. This can also be seen in the 1D profiles: filaments are generally cooler compared to their surroundings close to the cluster, whereas the opposite is true at large radial distances when comparing filaments with void-like regions. In the first case, filaments are embedded in cluster-like regions (protecting galaxies from the more hostile ICM in agreement with \citealt{Kotecha2022}), whereas in the second one they are surrounded by lower-density voids. If we consider the temperature at the filament spine, there is an evident increase towards the cluster centre which is not that pronounced at large distances, as we can deduce from the separation of the 1D profiles at $\theta = 0$. The abrupt increase of the temperature closer to the central regions could be the result of the blurred out contributions of individual temperature jumps in filaments when material flowing through them gets shock heated as it slows down \citep{Vurm2023}. These shocks could be related to some of the features observed in the Mach number panel of Fig.~\ref{fig:all1} near filament spines in the cluster outskirts.

The entropic function of the gas can be seen in the lower left panel. From this plot, it is clear that closer to the filament spine the entropy decreases to values similar to gas in \q{void} regions (i.e., $\theta \approx 90^{\circ}$ and $d_{\rm clus} \gtrsim 4 R_{200}$). This suggests that the hot gas inside filaments was subject to quasi-isoentropic processes responsible for increasing its temperature such as, adiabatic compression of the gas or simple compression, just like material in voids. Conversely, gas in filament outskirts shows larger entropy values suggesting that non-isoentropic processes, most likely accretion shocks, play a significant role to establish the observed profiles. This is confirmed in the 1D plots at large radial distances by the presence of a peak at an angular scale of $\theta \approx 10^{\circ}$. These results are in agreement with the work of \citet{Angelinelli2021}, where it is expected that gas in filament regions have relatively low entropy values, basing their filament identification tool on this premise. We also observe an increase of the entropy along the spine of the filament as we move towards the cluster centre in agreement with \citet{Vurm2023}. In Section~\ref{sec:phasediagram}, we will analyse the temporal evolution of stacked filamentary gas to shed light on its density and temperature evolution.

In the lower right panel we plot the fraction of gas particles belonging to the WHIM that we define as gas with temperatures in the range $10^{5} - 10^{7}\,$K and densities of $n_{\rm H} \lesssim 10^{-4}\,h^2\,{\rm cm}^{-3}$ \cite[e.g.][]{Martizzi2019,galarraga-espinosa2021}. It is evident that the WHIM fraction at $\theta \approx 0^{\circ}$ is close to $1$, proving again that filaments hold a large proportion of warm-hot intergalactic gas as seen in other works \citep[e.g.][]{galarraga-espinosa2021, Galarraga-Espinosa, Tuominen2021}. In fact, the 1D profiles clearly show that the fraction of WHIM gas monotonically grows towards the filament spine for all distances from the stacked cluster. This behaviour is different to the gas fraction profiles shown in \citet{galarraga-espinosa2021}, where they find a peak at a distance of about $1\,$Mpc from filaments. However, if we include material from haloes, we do obtain WHIM fraction profiles showing a peak at $\theta\approx10^{\circ}$ as it is shown in Fig.~\ref{fig:ApenStack0} (see below). At distances $2 R_{200}\lesssim d_{\rm clus} \lesssim 3 R_{200}$ the hot gas associated to the cluster lowers the WHIM fraction, except in the direction of the filament. Interestingly, this type of gas fills the cavity of hot gas already seen in the temperature maps along the spine direction indicating that the central cluster is constantly feed with fresh warm-hot intergalactic gas that can penetrate the hotter ICM. Contours indicate the gas density, calculated using randomly positioned particles to determine the volume of the complex shapes that result after removing haloes within filaments (see Section~\ref{sec:haloexcision}).

To assess the level of contamination provided by galactic haloes in our determination of filament properties, we have repeated the same analysis without excising haloes from the simulations. Generally speaking, we found similar results compared to the diffuse filamentary gas only, although the signal in this case is significantly noisier. We refer the reader to Figs.~\ref{fig:ApenStack0} and~\ref{fig:ApenStack1} in the Appendix~\ref{app:halo_cont}. Note that, when including haloes, the gas density is much higher compared to that of the diffuse material (see contour plots in the lower right panels of Figs.~\ref{fig:Stack_all} and~\ref{fig:ApenStack0}). Furthermore, the temperature is much noisier with hotspots distributed all around the cluster region. These \q{clumps} of gas host the warm circumgalactic medium (WCGM) of galaxies that, together with the WHIM, comprise all the warm gas available. This plot suggest that the WCGM of haloes in filaments could also play a significant role in future X-ray observations of filaments and their surrounding regions \citep{Angelinelli2021, Gouin2023}.

\subsubsection{Velocity fields}
\label{sec:vel_fields}

To determine the velocity components of a gas particle with respect to the filamentary structure, we define a coordinate system centred on the particle by taking (i) the salient radial unit vector from the centre of the cluster pointing in the direction of the particle, $u_{r}$, and (ii) the direction towards the nearest filament (as explained at the beginning of Section~\ref{sec:filamentgasprop}), $u_{\rm fil}$. Note that for the angular distance calculation, all filaments from the sample were considered. 

We first remove the radial component of the vector $u_{\rm fil}$, normalise it and flip its direction to get $u_{t}$:

\begin{equation}
   u_{t}\equiv -\frac{u_{\rm fil} - u_{r}(u_{r} \cdot u_{\rm fil}) }{|u_{\rm fil} - u_{r}(u_{r} \cdot u_{\rm fil})|}. 
\end{equation}

In this way, a vector that is pointing outwards from the location of the nearest filament is obtained. At same time, this unit vector is perpendicular to the radial direction from the cluster pointing to the particle. Then, we define $u_{p} = u_{t} \times u_{r}$. These three unit vectors are represented in green in the diagram of Fig.~\ref{fig:diagram}. After subtracting the velocity of the mass centre of the whole cluster, the velocity components are defined as the dot product of the velocity vectors and the unit vectors mentioned above.
This analysis of the velocity components exploits the roughly spherical symmetry of the cluster region. It allows us to easily separate the components influenced by the filament (tangential), and by just the central cluster (radial).

\begin{figure}
  \includegraphics[width=\columnwidth]{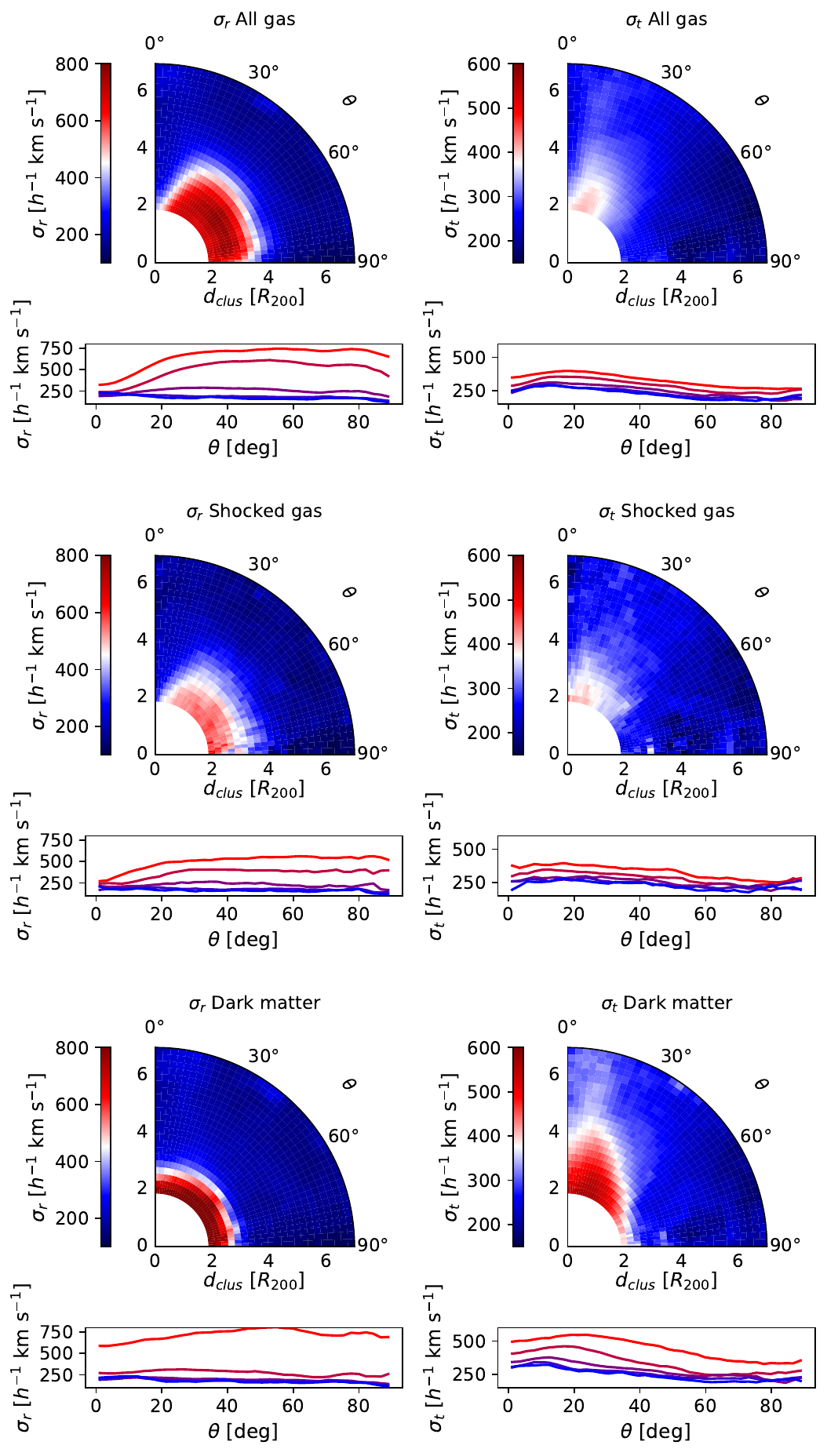}
  \caption[ArcDistanceDM]{Angular dependence of the velocity dispersion for all gas (upper row), shocked gas (middle row) and dark matter (lower row) for the radial and tangential components defined in the previous figure. Note the isotropy in the radial velocity dispersion of the dark matter component. The profiles below each panel show the plotted quantities along the separation angle $\theta$ at different radii from $d_{\rm clus} = 2R_{200}$ (red lines) to $7R_{200}$ (blue lines). 
}
  \label{fig:sigma}
\end{figure}

Fig.~\ref{fig:Stack_shocked} displays the components of the velocity vector for the stacked filament sample. On the top left we see the radial component of all gas, consisting of only negative velocities as material is generally infalling towards the cluster. For filamentary gas, velocities are maximally negative at $2 R_{200} \lesssim d_{\rm clus} \lesssim 3.5 R_{200}$. However, material at similar distances but outside filaments (i.e., $30^{\circ} \lesssim \theta \lesssim 80^{\circ}$) comprising both infalling and ejected gas is observed. These results are in accordance with the findings of \cite{Rost21}, \citet{Gouin2022} and \citet{Vurm2023}. In fact, some of this material might have been inside $R_{200}$ so that it could be considered ``backsplash'' gas, analogous to the backsplash galaxies studied in the literature \citep{Gill2005, Muriel2014, Haggar2020}, which refer to systems that have been inside the virial radius of a massive cluster in the past and might have been ``processed'' by it. This gas seems to avoid the filament direction when ejected, however this is not necessarily true in the case of haloes, since the presence of ``backsplash haloes'' in filaments has been recently shown by \cite{Kuchner2022}. This region around the filament axis and close to the $2R_{200}$ sphere is interesting because it contains gas funnelled by the filament onto the central cluster but, at the same time, haloes and galaxies travelling upstream.
The 1D profiles show that at large radial distances the infall of gas is slower inside the filament than in the surrounding region, probably owing to the gravitational pull of other cosmological structures connected to the filaments. These complex fluxes around filaments have also been extensively studied in works such as \citet{Codis2012} and \citet{Kraljic2018}. However, towards the centre we see a major increase in the infall velocity inside filaments compared to the surroundings. When taking into account only shocked particles (upper right panel of Fig.~\ref{fig:Stack_shocked}), the largest difference lies close to the cluster and in the range $30^{\circ} \lesssim \theta \lesssim 60^{\circ}$. The non-shocked gas seems to be going against the shocked one.

The tangential component of the velocity in the filament direction is shown in the lower row of Fig.~\ref{fig:Stack_shocked}, both for all and shocked gas respectively. As with the radial component, both values are negative while their magnitude greatly differ. This indicates that the material is infalling towards the spine in an approximately perpendicular direction to the local filament axis, at different speeds, with shocked gas moving faster than non-shocked material. Close to the spine, the tangential velocity drops to near zero values as approaching material is slowed down and swept by gas travelling through the inner regions towards the cluster. Another possible reason for the decrease of the tangential velocity could be a slight misalignment between the physical filament structure and the corresponding axis identified by the filament finder. \disperse\ detects the filament spine as a ridge that connects two critical points in the density field, not necessarily implying that the velocity field will be centred around this axis too. Further away, at angles $10^{\circ} \lesssim \theta \lesssim 60^{\circ}$, the gravitational attraction of these structures is more evident. Interestingly, at angular scales of around $90^{\circ}$, the gravitational influence of the filament vanishes, or is balanced out by the presence of another filament.

The 1D profiles also show a general infall of material at large distances of the stacked cluster towards the filament spine, however, close to the central region positive tangential velocities, peaking at $\theta \approx 15^{\circ}$, are seen. This is consistent with the presence of a shear resulting from the difference in velocity between gas inside filaments and hot gas surrounding the cluster.
When we compare the tangential component between shocked material and all gas particles clear differences can be observed. The most striking one is that shocked gas is infalling faster than average near the spine, suggesting that shocks are produced by gravitational infalling material that encounters gas flowing along the filament as it approaches the spine. Specifically, the shocked gas seems to be going against non-shocked material having larger negative velocities of about $-200\hkms$ in comparison to $-100\,\hkms$ for all gas. 

Finally, in Fig.~\ref{fig:sigma}, we show the velocity dispersion for all gas, shocked gas and dark matter for the radial and tangential components with respect to the filament spine. On the left-hand panel, the radial component is shown displaying a region located at $30^{\circ} \lesssim \theta \lesssim 70^{\circ}$ with a high velocity dispersion of about $600\,\hkms$ that correlates with the high-entropy distribution close to the cluster, similar in shape and size to the hot material seen in Fig.~\ref{fig:Stack_all}. This excludes the filament direction, where the gas seems to be flowing softly, thus providing a quiet environment for filament galaxies flowing towards the centre. As mentioned above, inner filament regions seem to be able to protect galaxies from the shocked material around the cluster, which is in agreement with the findings of \citet{Kotecha2022}. In the case of shocked gas, the velocity dispersion is lower with a dispersion of about $500\,\hkms$, although the peak is located at the same place as the average distribution. From the 1D plots it can be seen that the dispersion away from the cluster centre is more or less the same, whereas the difference between the spine and the surrounding region increases towards the centre.
The right-hand panel shows the dispersion of the tangential component. In this case, the maximum velocity is located around $2R_{200}$ at $\theta \approx 20^{\circ}$ and tends to have high values, except in the void-like region located at $3 < d_{\rm clus}/R_{200} < 7$ and $\theta \gtrsim 50^{\circ}$. This behaviour is consistent with the generation of turbulence in the transition layer between the filament and the hot gas associated with the cluster.

\subsection{Phase-diagram evolution}
\label{sec:phasediagram}

\begin{figure}
  \includegraphics[width=\columnwidth]{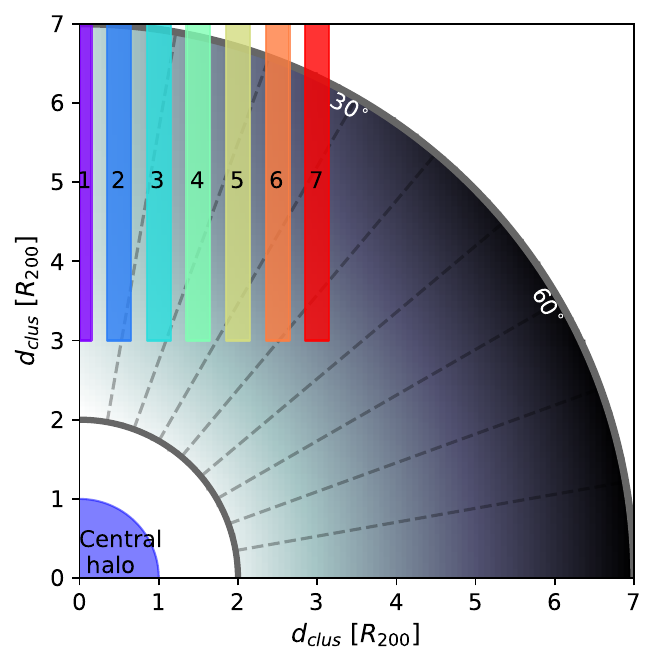}
  \caption{Location at $z=0$ of groups of gas particles surrounding filaments considered to trace back the evolution of different filament regions (see next figure).}
  \label{fig:stripes}
\end{figure}

\begin{figure}
  \includegraphics[width=\columnwidth]{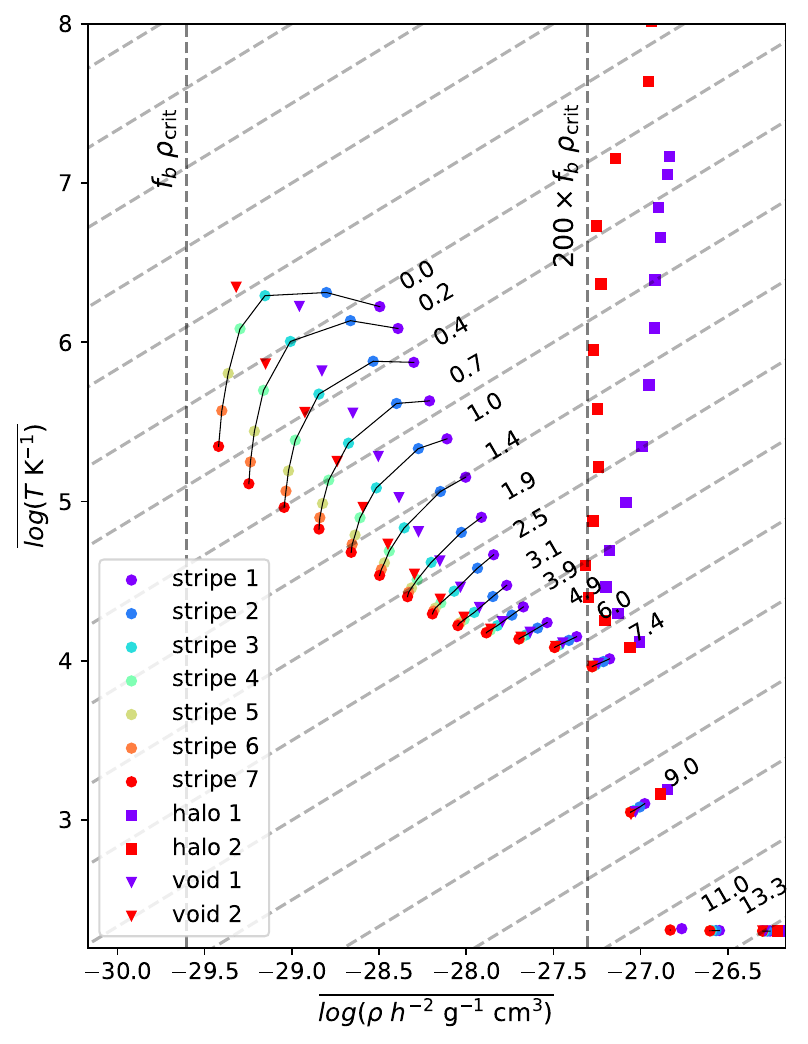}
  \caption{Phase-space diagram evolution for groups of gas particles that end up in different parts of filaments at $z=0$. The $x$-axis is the mean logarithm of density, whereas the $y$-axis is the mean logarithm of temperature. The numbers indicate the redshift for the case of the stripes. Purple-blue dots represent a group of particles located on/close to the spine of filaments at different distances from the central cluster, whereas green-red colours represent groups of particles located at filament outskirts, towards void regions. Iso-entropic trajectories are shown as grey dashed lines. For comparison, squares and triangles show the evolution of two halo samples and two void-like regions, respectively (see text).}
  \label{fig:phase}
\end{figure}

To study the nature of gas profiles in filaments we would like to understand the history of the accreted material. The timescales involved in the growth of filamentary structures, and the fact that their overdensities are at most of a few dozen times $\rho_{\rm crit}$, make them intermediate objects with internal structures highly contaminated by the presence of galaxy haloes. If we remove them to study only the diffuse filamentary material, a characteristic gas temperature profile that does not follow a pure adiabatic contraction is found (i.e.inner regions closer to the spine have lower temperatures than in the adjacent regions of filaments, as shown in Fig. \ref{fig:fit}). To understand the origin of this feature, we selected groups of particles, sorted by distance from the filamentary axis, forming a series of stripes around filaments belonging to the cluster sample at $z=0$ (see Fig. \ref{fig:stripes}). We tracked the temperature of these groups of particles back in time in the 129 snapshots and found that particles in the hotter stripes had a relatively sudden increase of their temperature, consistent with a shock, while the colder particles in the interior of filaments experienced a slower temperature increase. This is expected as the formation of accretion shocks around filaments is a natural outcome of cosmological structure formation. 
The resulting evolution is shown in Fig.~\ref{fig:phase}, where we plot the mean logarithmic temperature as a function of the mean logarithmic {\it physical} density of gas. The dashed lines represent the expected adiabatic relation between gas density and temperature, which should be followed by the phase-space trajectories in their linear regime of the evolution. In general, none of the regions follow a strict adiabatic evolution since all the dots travel roughly perpendicularly to the dashed lines. However, not all the regions move in the diagram in the same way. In particular, the cyan points (corresponding to regions located at a distance of about $\gtrsim R_{200}$ from the stacked filament spine) show the highest entropy at $z=0$. Note that purple dots represent the filament spine which, together with material in the outskirts (red dots), seem to have the lowest entropy at $z=0$, indicating that the inner regions are better protected by the shells surrounding filaments. Meanwhile, the intermediate layers are significantly affected by shock heating and the stirring of galaxy dynamics, in contrast to those farther away from shocks reaching lower entropy values. However, at earlier times up to $z = 1.9$, the trend is different since the purple dots have the highest entropy among all the filament regions.

For comparison, we trace back in time the evolution of all gas belonging to galaxy haloes within $R_{200}$ in simulated cluster region 1 at $z=0$ with a mass larger than $10^{11}\,h^{-1}\,$M$_{\odot}$ excluding the central halo. We refer to this choice as `halo 1' in Fig~\ref{fig:phase}. Additionally, all gas within $R_{200}$ of the central halo with $M_{200}(z=0)=2.62\times10^{15}\,h^{-1}
\,$M$_{\odot}$ is also shown (`halo 2'). As expected, the phase-diagram evolution of these haloes drastically differ from that of filaments. This is easily seen after virialisation, where they separate from the filament evolution, following their own path in phase-space as haloes collapse and accrete material. 

Contrary to haloes, both filaments and voids are non-fully virialised structures that evolve following cosmological cooling and expansion. Fig.~\ref{fig:phase} also shows two examples of void-like regions from simulated cluster region 1: (i) all gas outside $2R_{200}$ of the central cluster and outside filaments with $d_{\rm fil} > 3\,h^{-1}\,$Mpc, i.e. including both haloes and diffuse material (`void 1'), and (ii) excluding all haloes in the latter (`void 2'). In these two cases, the phase-space evolution of the material lies in a similar (non-virialiased) region to filamentary gas, with a tendency to cover the external parts of filaments. 

\begin{figure*}
\includegraphics[width=0.9\textwidth]{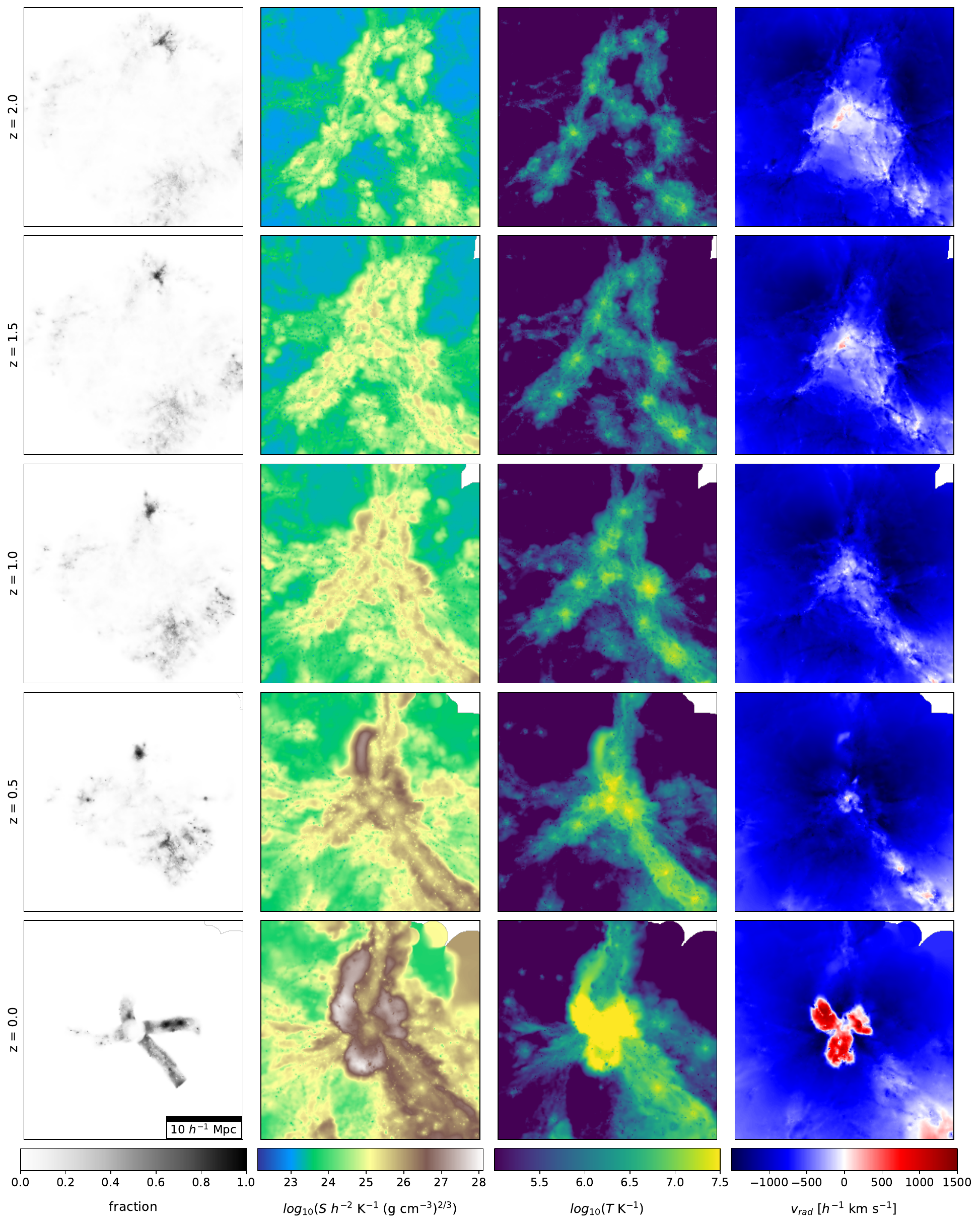}
  \caption{Time evolution of various physical quantities of a slice of $8\,h^{-1}\,$Mpc (comoving) thick in the simulated cluster region $14$. From left to right the columns display the fraction of material in the slice that will belong to a filament identified at $z = 0$, the entropic function of the gas, the gas temperature and the radial velocity towards the centre of the cluster.}
  \label{fig:evolution}
\end{figure*}

A visual impression of filament evolution can be seen directly from Fig.~\ref{fig:evolution}, where five different snapshots at  $z = 2, 1.5, 1, 0.5$ and $0$ for a $2\,h^{-1}\,$Mpc thick slice of cluster simulation $14$ are shown as an example. The first column from the left shows the fraction of gas that belongs to filaments at $z = 0$ outside the $R_{200}$ sphere of the corresponding cluster, where the assembly process of filaments can be clearly seen as they collapse in two directions, i.e. towards the cluster centre and perpendicularly. The second column displays the mean entropic function
of the gas. As shown above, high values owing to shock regions surrounding filaments at intermediate distances are seen at $z=0$, while the contrary is true near their core. However, this tendency is also observed  at earlier times: at $z>0$, {\it filamentary} structures with lower entropy values surrounded by high-entropy gas suggesting the presence of shocks are observed. The third column from the left shows the temperature distribution. In this case, high-temperature regions are correlated with high-entropy gas and, although less noticeably, filamentary regions also contain slightly cooler gas close to their spines in comparison to their surroundings.
Finally, the fourth column displays the radial component of gas velocity, where a burst of hot gas from the centre at $z = 0$ (not present in the previous snapshot) is seen. As already discussed in Fig.~\ref{fig:all1} these expanding lobes are produced as a result of a multiple-merger event at $z\approx 0.14$, highlighting the impact of mass assembly in the ICM.

\section{Discussion}
\label{sec:discus}
Although cosmological filaments around galaxy clusters are locations where specific gas properties could mildly affect galaxy properties compared to core regions of galaxy clusters, these environments are deeply linked to the cluster to which they are connected and are not free from complexities. In particular, in the outskirts of filaments, we have observed, confirming previous works, that there is a layer of shocked material that is heating up the gas which could have consequences for galaxy evolution. This layer could be related to high vorticity regions at the edges of filaments that, as  shown by \citet{Song2021}, could affect galaxy properties. Similar to the results of \citet{Klar-Mucket2012} for a smaller, idealised filament simulation, we observed a bump at a distance of about $1-2\,h^{-1}\,$Mpc from the filament axis in the mean temperature profile. However, in other works such as \citet{galarraga-espinosa2021} and \citet{Tuominen2021} the temperature profiles of different filament populations appear to be flatter from the core up to distances of about $2\,$Mpc, without a pronounced temperature increase. In our work, we also found that the average gas entropy is lower in the interior of filaments. To some extent, this observation agrees with the results of \citet{Angelinelli2021} as they are able to identify filaments by selecting low-entropy gas regions. This suggest that in the inner filament regions the gas is not heated by strong shocks in the same manner as in the filament outskirts. However, it is possible that inner filament regions have experienced shock formation as well, as the potential well of proto-filaments ending up in the observed structures today (although weaker than at the present time) were still capable of promoting shocks that would affect gas properties at the filament core. Furthermore, this material is being dragged rather calmly towards the centre of the cluster, resisting the action of the cluster's hot gas as it penetrates the inner regions of the cluster. This is in line with our previous work R21, \citet{Gouin2023} and also \citet{Vurm2023} who study in detail the formation of shocks in the gas ``rivers'' inside filaments when they bump into the hot atmosphere of the cluster. Some evidence of this can be seen in Fig.~\ref{fig:all1}. Our results suggest, in fact, that gas inside filaments was not heated up by strong shocks in the past but went through a process of pseudo-adiabatic compression as the radiative cooling inside filaments is not enough to explain the change in entropy. Conversely, the higher entropy of material surrounding filaments suggests that gas went through a series of stronger shocks, as discussed in the previous section.

After correlating the velocity fields with the entropic function, the WHIM distribution and the gas temperature, we infer that the filament structures collect their WHIM gas by efficiently dragging material from their surroundings \citep[as also found by][]{Tuominen2021,Galarraga-Espinosa}. Additionally, we do not observe an efficient isotropic collapse of WHIM material towards the central cluster. This can be seen when comparing the radial velocities of both all and shocked gas, where the higher infall velocities are preferentially linked to shock fronts going through the hot gas associated with the cluster in regions devoid of filaments. Additionally, we have checked for a dependence of the Mach number in filaments and spherical accretion shocks with mass and relaxation state of the cluster sample, but we have not found any statistically significant signal when distances are normalised to $R_{200}$. Similar to R21, our findings show that at distances close to $2R_{200}$ in the direction of filaments a ``cavity'' region forms where infall velocity is larger for gas than for dark matter. However, in the isotropic case, this relation is inverted, with gas moving slower than the dark matter towards the cluster centre. In general, we see that, relative to dark matter, baryons have a more negative infall towards the filament, which is also consistent with the fact that the tangential velocity dispersion is higher in the dark matter than the gas owing to its viscosity and pressure. When checking for the dispersion of the radial velocity component of the dark matter, we find no angular dependence, which is consistent with the velocity dispersion of the central galaxy cluster.

\section{Conclusions}

In this work, we used a set of galaxy cluster re-simulations belonging to the \threehun\ project aiming at reproducing cluster environments and their evolution within volumes of 15$\,h^{-1}\,$Mpc to study filamentary gas properties and accreting material towards filaments and clusters. To improve statistics, we stacked simulated clusters and measure average gas properties to produce density, temperature, pressure and entropy profiles and the Mach number distribution surrounding filaments at $z=0$. Additionally, we computed the average velocity field of baryons and dark matter in the sample and study the temporal evolution of filamentary gas in density-temperature phase-space. Our main results can be summarised as follows: 

\begin{itemize}
  
\item We confirm a different behaviour for baryons and dark matter in both filaments and galaxy cluster outskirts. We show that cosmic filaments are surrounded by shocks allowing the internal regions to efficiently transport fresh gas towards the cluster centre. The velocity dispersion of the gas suggests a laminar infall of material; meanwhile, the dark matter component shows a higher velocity dispersion, compatible with the properties of the central halo.\\ 

\item When considering filaments classified as {\it parallel} (i.e., those with a node in the central cluster) or {\it perpendicular} (i.e., those with a node in a parallel filament), there are no significant differences in the shock properties. However, there is a natural increase in filament thickness towards the main node, which is more evident for radial filaments towards the cluster centre.\\

\item We parameterise the shape of physical gas properties by fitting 1D profiles (in $R_{200}$ units of the central cluster) to our stacked `prominent' filament sample (see Table~\ref{tab:properties}), showing that entropy and temperature profiles display a bump at a characteristic scale length. Meanwhile, the density and pressure monotonically decrease with filament radius.\\

\item We demonstrate that the temperature profile of filaments is typically colder in the centre. Therefore, filaments naturally protect the inner regions, suggesting that the relatively pristine properties of baryons remain unchanged. In general, there should be no mixing, nor metals close to the spine (apart from those linked to galaxy haloes), whereas, around filaments, shocks rise the temperature and promote gas mixing through turbulence.\\

\item We show that diffuse gas in filaments does not escape from the cosmological expansion. However, this material does not follow a fully adiabatic evolution through time, showing the larger departures at intermediate distances. For our sample of simulated clusters at $z=0$ with a mean $R_{200}$ of $1.56\hMpc$, this translates in a distance of about $1-2\,h^{-1}\,$Mpc from the filament spine. These effects have to be taken into account when considering galaxy evolution in those regions.\\
  
\end{itemize}

Our work demonstrates that the spatial dependence of diffuse gas properties in filaments naturally define several environments that might differently affect galaxy evolution. Moreover, the effect of cosmological expansion in filament evolution shown in this work lead us to connect cosmological parameters and baryon physics. These aspects will be developed in future works.

\section*{Acknowledgments}
The authors acknowledge the anonymous referee for constructive comments that helped to improve this manuscript. 

SEN and FAS are members of the Carrera del Investigador Cient\'{\i}fico of CONICET. SEN acknowledges support by CONICET (project PIBAA R73734) and UBACyT 20020170100129BA. AR and FAS acknowledge support by ANPCyT (PICT-2016-4174) and SECyT-UNC (Consolidar 2018-2020). WC, AK and GY acknowledge Ministerio de  Ciencia e Innovaci\'on (Spain) for partial financial support under research grant PID2021-122603NB-C21.

This work has been made possible by \threehun\ collaboration\footnote{\url{https://www.the300-project.org}}. \threehun\ simulations used in this paper have been performed in the MareNostrum Supercomputer at the Barcelona Supercomputing Center, thanks to CPU time granted by the Red Espa\~nola de Supercomputaci\'on.
This work has received financial support from the European Union's Horizon 2020 Research and Innovation programme under the Marie Sk\l{}odowskaw-Curie grant agreement number 734374, i.e. the LACEGAL project\footnote{\url{https://cordis.europa.eu/project/rcn/207630\_en.html}}.
 For the purpose of open access, the author has applied a creative commons attribution (CC BY) to any author accepted manuscript version arising.
 
\section*{Data Availability}
The scripts and plots for this article will be shared on reasonable
request to the corresponding author.




\bibliographystyle{mnras}
\bibliography{biblio} 

\begin{thebibliography}{}
\makeatletter
\relax
\def\mn@urlcharsother{\let\do\@makeother \do\$\do\&\do\#\do\^\do\_\do\%\do\~}
\def\mn@doi{\begingroup\mn@urlcharsother \@ifnextchar [ {\mn@doi@}
  {\mn@doi@[]}}
\def\mn@doi@[#1]#2{\def\@tempa{#1}\ifx\@tempa\@empty \href
  {http://dx.doi.org/#2} {doi:#2}\else \href {http://dx.doi.org/#2} {#1}\fi
  \endgroup}
\def\mn@eprint#1#2{\mn@eprint@#1:#2::\@nil}
\def\mn@eprint@arXiv#1{\href {http://arxiv.org/abs/#1} {{\tt arXiv:#1}}}
\def\mn@eprint@dblp#1{\href {http://dblp.uni-trier.de/rec/bibtex/#1.xml}
  {dblp:#1}}
\def\mn@eprint@#1:#2:#3:#4\@nil{\def\@tempa {#1}\def\@tempb {#2}\def\@tempc
  {#3}\ifx \@tempc \@empty \let \@tempc \@tempb \let \@tempb \@tempa \fi \ifx
  \@tempb \@empty \def\@tempb {arXiv}\fi \@ifundefined
  {mn@eprint@\@tempb}{\@tempb:\@tempc}{\expandafter \expandafter \csname
  mn@eprint@\@tempb\endcsname \expandafter{\@tempc}}}

\bibitem[\protect\citeauthoryear{{Akamatsu} et~al.,}{{Akamatsu}
  et~al.}{2017}]{Akamatsu2017}
{Akamatsu} H.,  et~al., 2017, \mn@doi [\aap] {10.1051/0004-6361/201730497},
  \href {https://ui.adsabs.harvard.edu/abs/2017A&A...606A...1A} {606, A1}

\bibitem[\protect\citeauthoryear{{Angelinelli}, {Ettori}, {Vazza}  \&
  {Jones}}{{Angelinelli} et~al.}{2021}]{Angelinelli2021}
{Angelinelli} M.,  {Ettori} S.,  {Vazza} F.,   {Jones} T.~W.,  2021, \mn@doi
  [\aap] {10.1051/0004-6361/202140471}, \href
  {https://ui.adsabs.harvard.edu/abs/2021A&A...653A.171A} {653, A171}

\bibitem[\protect\citeauthoryear{{Araya-Melo}, {Arag{\'o}n-Calvo},
  {Br{\"u}ggen}  \& {Hoeft}}{{Araya-Melo} et~al.}{2012}]{ArayaMelo12}
{Araya-Melo} P.~A.,  {Arag{\'o}n-Calvo} M.~A.,  {Br{\"u}ggen} M.,   {Hoeft} M.,
   2012, \mn@doi [\mnras] {10.1111/j.1365-2966.2012.21042.x}, \href
  {https://ui.adsabs.harvard.edu/abs/2012MNRAS.423.2325A} {423, 2325}

\bibitem[\protect\citeauthoryear{{Arthur} et~al.,}{{Arthur}
  et~al.}{2019}]{Arthur2019}
{Arthur} J.,  et~al., 2019, \mn@doi [\mnras] {10.1093/mnras/stz212}, \href
  {https://ui.adsabs.harvard.edu/abs/2019MNRAS.484.3968A} {484, 3968}

\bibitem[\protect\citeauthoryear{{Bacon} et~al.,}{{Bacon}
  et~al.}{2021}]{Bacon2021}
{Bacon} R.,  et~al., 2021, \mn@doi [\aap] {10.1051/0004-6361/202039887}, \href
  {https://ui.adsabs.harvard.edu/abs/2021A&A...647A.107B} {647, A107}

\bibitem[\protect\citeauthoryear{{Beck} et~al.,}{{Beck}
  et~al.}{2016}]{Beck2016}
{Beck} A.~M.,  et~al., 2016, \mn@doi [\mnras] {10.1093/mnras/stv2443}, \href
  {https://ui.adsabs.harvard.edu/abs/2016MNRAS.455.2110B} {455, 2110}

\bibitem[\protect\citeauthoryear{{Behroozi}, {Wechsler}  \& {Wu}}{{Behroozi}
  et~al.}{2013}]{rockstar}
{Behroozi} P.~S.,  {Wechsler} R.~H.,   {Wu} H.-Y.,  2013, \mn@doi [\apj]
  {10.1088/0004-637X/762/2/109}, \href
  {https://ui.adsabs.harvard.edu/abs/2013ApJ...762..109B} {762, 109}

\bibitem[\protect\citeauthoryear{{Benitez-Llambay}}{{Benitez-Llambay}}{2015}]{sph-viewer}
{Benitez-Llambay} A.,  2015, {Py-Sphviewer: Py-Sphviewer V1.0.0},
  \mn@doi{10.5281/zenodo.21703}

\bibitem[\protect\citeauthoryear{{Bond}, {Kofman}  \& {Pogosyan}}{{Bond}
  et~al.}{1996}]{Bond1996}
{Bond} J.~R.,  {Kofman} L.,   {Pogosyan} D.,  1996, \mn@doi [\nat]
  {10.1038/380603a0}, \href
  {https://ui.adsabs.harvard.edu/abs/1996Natur.380..603B} {380, 603}

\bibitem[\protect\citeauthoryear{{Bonjean}, {Aghanim}, {Salom{\'e}}, {Douspis}
  \& {Beelen}}{{Bonjean} et~al.}{2018}]{Bonjean2018}
{Bonjean} V.,  {Aghanim} N.,  {Salom{\'e}} P.,  {Douspis} M.,   {Beelen} A.,
  2018, \mn@doi [\aap] {10.1051/0004-6361/201731699}, \href
  {https://ui.adsabs.harvard.edu/abs/2018A&A...609A..49B} {609, A49}

\bibitem[\protect\citeauthoryear{{Cautun}, {van de Weygaert}, {Jones}  \&
  {Frenk}}{{Cautun} et~al.}{2014}]{Cautun2014}
{Cautun} M.,  {van de Weygaert} R.,  {Jones} B. J.~T.,   {Frenk} C.~S.,  2014,
  \mn@doi [\mnras] {10.1093/mnras/stu768}, \href
  {https://ui.adsabs.harvard.edu/abs/2014MNRAS.441.2923C} {441, 2923}

\bibitem[\protect\citeauthoryear{{Cen} \& {Ostriker}}{{Cen} \&
  {Ostriker}}{1999}]{Cen99}
{Cen} R.,  {Ostriker} J.~P.,  1999, \mn@doi [\apj] {10.1086/306949}, \href
  {https://ui.adsabs.harvard.edu/abs/1999ApJ...514....1C} {514, 1}

\bibitem[\protect\citeauthoryear{{Codis}, {Pichon}, {Devriendt}, {Slyz},
  {Pogosyan}, {Dubois}  \& {Sousbie}}{{Codis} et~al.}{2012}]{Codis2012}
{Codis} S.,  {Pichon} C.,  {Devriendt} J.,  {Slyz} A.,  {Pogosyan} D.,
  {Dubois} Y.,   {Sousbie} T.,  2012, \mn@doi [\mnras]
  {10.1111/j.1365-2966.2012.21636.x}, \href
  {https://ui.adsabs.harvard.edu/abs/2012MNRAS.427.3320C} {427, 3320}

\bibitem[\protect\citeauthoryear{{Codis}, {Pogosyan}  \& {Pichon}}{{Codis}
  et~al.}{2018}]{Codis2018}
{Codis} S.,  {Pogosyan} D.,   {Pichon} C.,  2018, \mn@doi [\mnras]
  {10.1093/mnras/sty1643}, \href
  {https://ui.adsabs.harvard.edu/abs/2018MNRAS.479..973C} {479, 973}

\bibitem[\protect\citeauthoryear{{Colberg}, {Krughoff}  \&
  {Connolly}}{{Colberg} et~al.}{2005}]{Colberg2005}
{Colberg} J.~M.,  {Krughoff} K.~S.,   {Connolly} A.~J.,  2005, \mn@doi [\mnras]
  {10.1111/j.1365-2966.2005.08897.x}, \href
  {https://ui.adsabs.harvard.edu/abs/2005MNRAS.359..272C} {359, 272}

\bibitem[\protect\citeauthoryear{{Colless} et~al.,}{{Colless}
  et~al.}{2003}]{Colless2003}
{Colless} M.,  et~al., 2003, \mn@doi [arXiv e-prints]
  {10.48550/arXiv.astro-ph/0306581}, \href
  {https://ui.adsabs.harvard.edu/abs/2003astro.ph..6581C} {pp
  astro--ph/0306581}

\bibitem[\protect\citeauthoryear{{Cui} et~al.,}{{Cui} et~al.}{2018}]{Cui2018}
{Cui} W.,  et~al., 2018, \mn@doi [\mnras] {10.1093/mnras/sty2111}, \href
  {https://ui.adsabs.harvard.edu/abs/2018MNRAS.480.2898C} {480, 2898}

\bibitem[\protect\citeauthoryear{{Dekel} et~al.,}{{Dekel}
  et~al.}{2009}]{Dekel2009}
{Dekel} A.,  et~al., 2009, \mn@doi [\nat] {10.1038/nature07648}, \href
  {https://ui.adsabs.harvard.edu/abs/2009Natur.457..451D} {457, 451}

\bibitem[\protect\citeauthoryear{{Gal{\'a}rraga-Espinosa}, {Aghanim}, {Langer},
  {Gouin}  \& {Malavasi}}{{Gal{\'a}rraga-Espinosa}
  et~al.}{2020}]{Galarraga-Espinosa2020}
{Gal{\'a}rraga-Espinosa} D.,  {Aghanim} N.,  {Langer} M.,  {Gouin} C.,
  {Malavasi} N.,  2020, \mn@doi [\aap] {10.1051/0004-6361/202037986}, \href
  {https://ui.adsabs.harvard.edu/abs/2020A&A...641A.173G} {641, A173}

\bibitem[\protect\citeauthoryear{{Gal{\'a}rraga-Espinosa}, {Aghanim}, {Langer}
  \& {Tanimura}}{{Gal{\'a}rraga-Espinosa}
  et~al.}{2021}]{galarraga-espinosa2021}
{Gal{\'a}rraga-Espinosa} D.,  {Aghanim} N.,  {Langer} M.,   {Tanimura} H.,
  2021, \mn@doi [\aap] {10.1051/0004-6361/202039781}, \href
  {https://ui.adsabs.harvard.edu/abs/2021A&A...649A.117G} {649, A117}

\bibitem[\protect\citeauthoryear{{Gal{\'a}rraga-Espinosa}, {Langer}  \&
  {Aghanim}}{{Gal{\'a}rraga-Espinosa} et~al.}{2022}]{Galarraga-Espinosa}
{Gal{\'a}rraga-Espinosa} D.,  {Langer} M.,   {Aghanim} N.,  2022, \mn@doi
  [\aap] {10.1051/0004-6361/202141974}, \href
  {https://ui.adsabs.harvard.edu/abs/2022A&A...661A.115G} {661, A115}

\bibitem[\protect\citeauthoryear{{Gill}, {Knebe}  \& {Gibson}}{{Gill}
  et~al.}{2005}]{Gill2005}
{Gill} S. P.~D.,  {Knebe} A.,   {Gibson} B.~K.,  2005, \mn@doi [\mnras]
  {10.1111/j.1365-2966.2004.08562.x}, \href
  {https://ui.adsabs.harvard.edu/abs/2005MNRAS.356.1327G} {356, 1327}

\bibitem[\protect\citeauthoryear{{Gouin}, {Gallo}  \& {Aghanim}}{{Gouin}
  et~al.}{2022}]{Gouin2022}
{Gouin} C.,  {Gallo} S.,   {Aghanim} N.,  2022, \mn@doi [\aap]
  {10.1051/0004-6361/202243032}, \href
  {https://ui.adsabs.harvard.edu/abs/2022A&A...664A.198G} {664, A198}

\bibitem[\protect\citeauthoryear{{Gouin}, {Bonamente}, {Galarraga-Espinosa},
  {Walker}  \& {Mirakhor}}{{Gouin} et~al.}{2023}]{Gouin2023}
{Gouin} C.,  {Bonamente} M.,  {Galarraga-Espinosa} D.,  {Walker} S.,
  {Mirakhor} M.,  2023, \mn@doi [arXiv e-prints] {10.48550/arXiv.2306.04694},
  \href {https://ui.adsabs.harvard.edu/abs/2023arXiv230604694G} {p.
  arXiv:2306.04694}

\bibitem[\protect\citeauthoryear{{Haggar}, {Gray}, {Pearce}, {Knebe}, {Cui},
  {Mostoghiu}  \& {Yepes}}{{Haggar} et~al.}{2020}]{Haggar2020}
{Haggar} R.,  {Gray} M.~E.,  {Pearce} F.~R.,  {Knebe} A.,  {Cui} W.,
  {Mostoghiu} R.,   {Yepes} G.,  2020, \mn@doi [\mnras]
  {10.1093/mnras/staa273}, \href
  {https://ui.adsabs.harvard.edu/abs/2020MNRAS.492.6074H} {492, 6074}

\bibitem[\protect\citeauthoryear{{Haines} et~al.,}{{Haines}
  et~al.}{2015}]{Heines2015}
{Haines} C.~P.,  et~al., 2015, \mn@doi [\apj] {10.1088/0004-637X/806/1/101},
  \href {https://ui.adsabs.harvard.edu/abs/2015ApJ...806..101H} {806, 101}

\bibitem[\protect\citeauthoryear{{Haines} et~al.,}{{Haines}
  et~al.}{2017}]{Heines2017}
{Haines} C.~P.,  et~al., 2017, \mn@doi [\aap] {10.1051/0004-6361/201630118},
  \href {https://ui.adsabs.harvard.edu/abs/2017A&A...605A...4H} {605, A4}

\bibitem[\protect\citeauthoryear{{Hoeft}, {Nuza}, {Gottl{\"o}ber}, {van
  Weeren}, {R{\"o}ttgering}  \& {Br{\"u}ggen}}{{Hoeft} et~al.}{2011}]{Hoeft11}
{Hoeft} M.,  {Nuza} S.~E.,  {Gottl{\"o}ber} S.,  {van Weeren} R.~J.,
  {R{\"o}ttgering} H.~J.~A.,   {Br{\"u}ggen} M.,  2011, \mn@doi [Journal of
  Astrophysics and Astronomy] {10.1007/s12036-011-9127-z}, \href
  {https://ui.adsabs.harvard.edu/abs/2011JApA...32..509H} {32, 509}

\bibitem[\protect\citeauthoryear{{Klar} \& {M{\"u}cket}}{{Klar} \&
  {M{\"u}cket}}{2012}]{Klar-Mucket2012}
{Klar} J.~S.,  {M{\"u}cket} J.~P.,  2012, \mn@doi [\mnras]
  {10.1111/j.1365-2966.2012.20877.x}, \href
  {https://ui.adsabs.harvard.edu/abs/2012MNRAS.423..304K} {423, 304}

\bibitem[\protect\citeauthoryear{{Klypin}, {Yepes}, {Gottl{\"o}ber}, {Prada}
  \& {He{\ss}}}{{Klypin} et~al.}{2016}]{multidark}
{Klypin} A.,  {Yepes} G.,  {Gottl{\"o}ber} S.,  {Prada} F.,   {He{\ss}} S.,
  2016, \mn@doi [\mnras] {10.1093/mnras/stw248}, \href
  {https://ui.adsabs.harvard.edu/abs/2016MNRAS.457.4340K} {457, 4340}

\bibitem[\protect\citeauthoryear{{Knebe}, {Gill}, {Gibson}, {Lewis}, {Ibata}
  \& {Dopita}}{{Knebe} et~al.}{2004}]{Knebe2004}
{Knebe} A.,  {Gill} S. P.~D.,  {Gibson} B.~K.,  {Lewis} G.~F.,  {Ibata} R.~A.,
   {Dopita} M.~A.,  2004, \mn@doi [\apj] {10.1086/381306}, \href
  {https://ui.adsabs.harvard.edu/abs/2004ApJ...603....7K} {603, 7}

\bibitem[\protect\citeauthoryear{{Knollmann} \& {Knebe}}{{Knollmann} \&
  {Knebe}}{2009}]{AHF}
{Knollmann} S.~R.,  {Knebe} A.,  2009, \mn@doi [\apjs]
  {10.1088/0067-0049/182/2/608}, \href
  {https://ui.adsabs.harvard.edu/abs/2009ApJS..182..608K} {182, 608}

\bibitem[\protect\citeauthoryear{{Kotecha} et~al.,}{{Kotecha}
  et~al.}{2022}]{Kotecha2022}
{Kotecha} S.,  et~al., 2022, \mn@doi [\mnras] {10.1093/mnras/stac300}, \href
  {https://ui.adsabs.harvard.edu/abs/2022MNRAS.512..926K} {512, 926}

\bibitem[\protect\citeauthoryear{{Kraljic} et~al.,}{{Kraljic}
  et~al.}{2018}]{Kraljic2018}
{Kraljic} K.,  et~al., 2018, \mn@doi [\mnras] {10.1093/mnras/stx2638}, \href
  {https://ui.adsabs.harvard.edu/abs/2018MNRAS.474..547K} {474, 547}

\bibitem[\protect\citeauthoryear{{Kuchner} et~al.,}{{Kuchner}
  et~al.}{2022}]{Kuchner2022}
{Kuchner} U.,  et~al., 2022, \mn@doi [\mnras] {10.1093/mnras/stab3419}, \href
  {https://ui.adsabs.harvard.edu/abs/2022MNRAS.510..581K} {510, 581}

\bibitem[\protect\citeauthoryear{{Landau} \& {Lifshitz}}{{Landau} \&
  {Lifshitz}}{1959}]{Landau59}
{Landau} L.~D.,  {Lifshitz} E.~M.,  1959, {Fluid mechanics}

\bibitem[\protect\citeauthoryear{{Mart{\'\i}nez}, {Muriel}  \&
  {Coenda}}{{Mart{\'\i}nez} et~al.}{2016}]{Martinez2016}
{Mart{\'\i}nez} H.~J.,  {Muriel} H.,   {Coenda} V.,  2016, \mn@doi [\mnras]
  {10.1093/mnras/stv2295}, \href
  {https://ui.adsabs.harvard.edu/abs/2016MNRAS.455..127M} {455, 127}

\bibitem[\protect\citeauthoryear{{Martizzi} et~al.,}{{Martizzi}
  et~al.}{2019}]{Martizzi2019}
{Martizzi} D.,  et~al., 2019, \mn@doi [\mnras] {10.1093/mnras/stz1106}, \href
  {https://ui.adsabs.harvard.edu/abs/2019MNRAS.486.3766M} {486, 3766}

\bibitem[\protect\citeauthoryear{{Miniati}, {Ryu}, {Kang}, {Jones}, {Cen}  \&
  {Ostriker}}{{Miniati} et~al.}{2000}]{Miniati00}
{Miniati} F.,  {Ryu} D.,  {Kang} H.,  {Jones} T.~W.,  {Cen} R.,   {Ostriker}
  J.~P.,  2000, \mn@doi [\apj] {10.1086/317027}, \href
  {https://ui.adsabs.harvard.edu/abs/2000ApJ...542..608M} {542, 608}

\bibitem[\protect\citeauthoryear{{Murante}, {Monaco}, {Giovalli}, {Borgani}  \&
  {Diaferio}}{{Murante} et~al.}{2010}]{Murante2010}
{Murante} G.,  {Monaco} P.,  {Giovalli} M.,  {Borgani} S.,   {Diaferio} A.,
  2010, \mn@doi [\mnras] {10.1111/j.1365-2966.2010.16567.x}, \href
  {https://ui.adsabs.harvard.edu/abs/2010MNRAS.405.1491M} {405, 1491}

\bibitem[\protect\citeauthoryear{{Muriel} \& {Coenda}}{{Muriel} \&
  {Coenda}}{2014}]{Muriel2014}
{Muriel} H.,  {Coenda} V.,  2014, \mn@doi [\aap] {10.1051/0004-6361/201322033},
  \href {https://ui.adsabs.harvard.edu/abs/2014A&A...564A..85M} {564, A85}

\bibitem[\protect\citeauthoryear{{Nelson} et~al.,}{{Nelson}
  et~al.}{2019}]{Nelson2019}
{Nelson} D.,  et~al., 2019, \mn@doi [Computational Astrophysics and Cosmology]
  {10.1186/s40668-019-0028-x}, \href
  {https://ui.adsabs.harvard.edu/abs/2019ComAC...6....2N} {6, 2}

\bibitem[\protect\citeauthoryear{{Nuza}}{{Nuza}}{2023}]{Nuza23}
{Nuza} S.~E.,  2023, Bolet\'{\i}n de la Asociaci\'on Argentina de
  Astronom\'{\i}a, \href
  {https://ui.adsabs.harvard.edu/abs/2023BAAA...64..166N} {64, 166}

\bibitem[\protect\citeauthoryear{{Nuza}, {Hoeft}, {van Weeren}, {Gottl{\"o}ber}
   \& {Yepes}}{{Nuza} et~al.}{2012}]{Nuza12}
{Nuza} S.~E.,  {Hoeft} M.,  {van Weeren} R.~J.,  {Gottl{\"o}ber} S.,   {Yepes}
  G.,  2012, \mn@doi [\mnras] {10.1111/j.1365-2966.2011.20118.x}, \href
  {https://ui.adsabs.harvard.edu/abs/2012MNRAS.420.2006N} {420, 2006}

\bibitem[\protect\citeauthoryear{{Nuza}, {Gelszinnis}, {Hoeft}  \&
  {Yepes}}{{Nuza} et~al.}{2017}]{Nuza17}
{Nuza} S.~E.,  {Gelszinnis} J.,  {Hoeft} M.,   {Yepes} G.,  2017, \mn@doi
  [\mnras] {10.1093/mnras/stx1109}, \href
  {https://ui.adsabs.harvard.edu/abs/2017MNRAS.470..240N} {470, 240}

\bibitem[\protect\citeauthoryear{{Peacock}}{{Peacock}}{1999}]{Peacock1999}
{Peacock} J.~A.,  1999, {Cosmological Physics}

\bibitem[\protect\citeauthoryear{{Peebles}}{{Peebles}}{1980}]{Peebles1980}
{Peebles} P.~J.~E.,  1980, {The large-scale structure of the universe}

\bibitem[\protect\citeauthoryear{{Planck Collaboration} et~al.,}{{Planck
  Collaboration} et~al.}{2020}]{Planck2018}
{Planck Collaboration} et~al., 2020, \mn@doi [\aap]
  {10.1051/0004-6361/201833910}, \href
  {https://ui.adsabs.harvard.edu/abs/2020A&A...641A...6P} {641, A6}

\bibitem[\protect\citeauthoryear{{Porter}, {Raychaudhury}, {Pimbblet}  \&
  {Drinkwater}}{{Porter} et~al.}{2008}]{Porter2008}
{Porter} S.~C.,  {Raychaudhury} S.,  {Pimbblet} K.~A.,   {Drinkwater} M.~J.,
  2008, \mn@doi [\mnras] {10.1111/j.1365-2966.2008.13388.x}, \href
  {https://ui.adsabs.harvard.edu/abs/2008MNRAS.388.1152P} {388, 1152}

\bibitem[\protect\citeauthoryear{{Power} et~al.,}{{Power}
  et~al.}{2020}]{Power2020}
{Power} C.,  et~al., 2020, \mn@doi [\mnras] {10.1093/mnras/stz3176}, \href
  {https://ui.adsabs.harvard.edu/abs/2020MNRAS.491.3923P} {491, 3923}

\bibitem[\protect\citeauthoryear{{Rams{\o}y}, {Slyz}, {Devriendt}, {Laigle}  \&
  {Dubois}}{{Rams{\o}y} et~al.}{2021}]{Ramsoy2021}
{Rams{\o}y} M.,  {Slyz} A.,  {Devriendt} J.,  {Laigle} C.,   {Dubois} Y.,
  2021, \mn@doi [\mnras] {10.1093/mnras/stab015}, \href
  {https://ui.adsabs.harvard.edu/abs/2021MNRAS.502..351R} {502, 351}

\bibitem[\protect\citeauthoryear{{Rasia} et~al.,}{{Rasia}
  et~al.}{2015}]{Rasia2015}
{Rasia} E.,  et~al., 2015, \mn@doi [\apjl] {10.1088/2041-8205/813/1/L17}, \href
  {https://ui.adsabs.harvard.edu/abs/2015ApJ...813L..17R} {813, L17}

\bibitem[\protect\citeauthoryear{{Rost} et~al.,}{{Rost} et~al.}{2021}]{Rost21}
{Rost} A.,  et~al., 2021, \mn@doi [\mnras] {10.1093/mnras/staa3792}, \href
  {https://ui.adsabs.harvard.edu/abs/2021MNRAS.502..714R} {502, 714}

\bibitem[\protect\citeauthoryear{{Song} et~al.,}{{Song}
  et~al.}{2021}]{Song2021}
{Song} H.,  et~al., 2021, \mn@doi [\mnras] {10.1093/mnras/staa3981}, \href
  {https://ui.adsabs.harvard.edu/abs/2021MNRAS.501.4635S} {501, 4635}

\bibitem[\protect\citeauthoryear{{Sousbie}}{{Sousbie}}{2011}]{Sousbie2011}
{Sousbie} T.,  2011, \mn@doi [\mnras] {10.1111/j.1365-2966.2011.18394.x}, \href
  {https://ui.adsabs.harvard.edu/abs/2011MNRAS.414..350S} {414, 350}

\bibitem[\protect\citeauthoryear{{Springel}}{{Springel}}{2005}]{Springel2005}
{Springel} V.,  2005, \mn@doi [\mnras] {10.1111/j.1365-2966.2005.09655.x},
  \href {https://ui.adsabs.harvard.edu/abs/2005MNRAS.364.1105S} {364, 1105}

\bibitem[\protect\citeauthoryear{{Sunyaev} \& {Zeldovich}}{{Sunyaev} \&
  {Zeldovich}}{1972}]{SunZel72}
{Sunyaev} R.~A.,  {Zeldovich} Y.~B.,  1972, \aap, \href
  {https://ui.adsabs.harvard.edu/abs/1972A&A....20..189S} {20, 189}

\bibitem[\protect\citeauthoryear{{Tanimura}, {Aghanim}, {Douspis}, {Beelen}  \&
  {Bonjean}}{{Tanimura} et~al.}{2019}]{Tanimura2019}
{Tanimura} H.,  {Aghanim} N.,  {Douspis} M.,  {Beelen} A.,   {Bonjean} V.,
  2019, \mn@doi [\aap] {10.1051/0004-6361/201833413}, \href
  {https://ui.adsabs.harvard.edu/abs/2019A&A...625A..67T} {625, A67}

\bibitem[\protect\citeauthoryear{{Tittley} \& {Henriksen}}{{Tittley} \&
  {Henriksen}}{2001}]{Tittley2001}
{Tittley} E.~R.,  {Henriksen} M.,  2001, \mn@doi [\apj] {10.1086/323955}, \href
  {https://ui.adsabs.harvard.edu/abs/2001ApJ...563..673T} {563, 673}

\bibitem[\protect\citeauthoryear{{Tuominen} et~al.,}{{Tuominen}
  et~al.}{2021}]{Tuominen2021}
{Tuominen} T.,  et~al., 2021, \mn@doi [\aap] {10.1051/0004-6361/202039221},
  \href {https://ui.adsabs.harvard.edu/abs/2021A&A...646A.156T} {646, A156}

\bibitem[\protect\citeauthoryear{{Umehata} et~al.,}{{Umehata}
  et~al.}{2019}]{Umehata2019}
{Umehata} H.,  et~al., 2019, \mn@doi [Science] {10.1126/science.aaw5949}, \href
  {https://ui.adsabs.harvard.edu/abs/2019Sci...366...97U} {366, 97}

\bibitem[\protect\citeauthoryear{{Vernstrom}, {Heald}, {Vazza}, {Galvin},
  {West}, {Locatelli}, {Fornengo}  \& {Pinetti}}{{Vernstrom}
  et~al.}{2021}]{Vernstrom2021}
{Vernstrom} T.,  {Heald} G.,  {Vazza} F.,  {Galvin} T.~J.,  {West} J.~L.,
  {Locatelli} N.,  {Fornengo} N.,   {Pinetti} E.,  2021, \mn@doi [\mnras]
  {10.1093/mnras/stab1301}, \href
  {https://ui.adsabs.harvard.edu/abs/2021MNRAS.505.4178V} {505, 4178}

\bibitem[\protect\citeauthoryear{{Vurm}, {Nevalainen}, {Hong}, {Bah{\'e}},
  {Dalla Vecchia}  \& {Hein{\"a}m{\"a}ki}}{{Vurm} et~al.}{2023}]{Vurm2023}
{Vurm} I.,  {Nevalainen} J.,  {Hong} S.~E.,  {Bah{\'e}} Y.~M.,  {Dalla Vecchia}
  C.,   {Hein{\"a}m{\"a}ki} P.,  2023, \mn@doi [\aap]
  {10.1051/0004-6361/202243904}, \href
  {https://ui.adsabs.harvard.edu/abs/2023A&A...673A..62V} {673, A62}

\bibitem[\protect\citeauthoryear{{Werner}, {Finoguenov}, {Kaastra},
  {Simionescu}, {Dietrich}, {Vink}  \& {B{\"o}hringer}}{{Werner}
  et~al.}{2008}]{Werner2008}
{Werner} N.,  {Finoguenov} A.,  {Kaastra} J.~S.,  {Simionescu} A.,  {Dietrich}
  J.~P.,  {Vink} J.,   {B{\"o}hringer} H.,  2008, \mn@doi [\aap]
  {10.1051/0004-6361:200809599}, \href
  {https://ui.adsabs.harvard.edu/abs/2008A&A...482L..29W} {482, L29}

\bibitem[\protect\citeauthoryear{{Wetzel}, {Tinker}, {Conroy}  \& {van den
  Bosch}}{{Wetzel} et~al.}{2013}]{Wetzel2013}
{Wetzel} A.~R.,  {Tinker} J.~L.,  {Conroy} C.,   {van den Bosch} F.~C.,  2013,
  \mn@doi [\mnras] {10.1093/mnras/stt469}, \href
  {https://ui.adsabs.harvard.edu/abs/2013MNRAS.432..336W} {432, 336}

\bibitem[\protect\citeauthoryear{{Zel'dovich}}{{Zel'dovich}}{1970}]{Zeldovich1970}
{Zel'dovich} Y.~B.,  1970, \aap, \href
  {https://ui.adsabs.harvard.edu/abs/1970A&A.....5...84Z} {5, 84}

\bibitem[\protect\citeauthoryear{{de Lapparent}, {Geller}  \& {Huchra}}{{de
  Lapparent} et~al.}{1986}]{DeLapparent1986}
{de Lapparent} V.,  {Geller} M.~J.,   {Huchra} J.~P.,  1986, \mn@doi [\apjl]
  {10.1086/184625}, \href
  {https://ui.adsabs.harvard.edu/abs/1986ApJ...302L...1D} {302, L1}

\bibitem[\protect\citeauthoryear{{van Haarlem} \& {van de Weygaert}}{{van
  Haarlem} \& {van de Weygaert}}{1993}]{vanHaarlem1993}
{van Haarlem} M.,  {van de Weygaert} R.,  1993, \mn@doi [\apj]
  {10.1086/173416}, \href
  {https://ui.adsabs.harvard.edu/abs/1993ApJ...418..544V} {418, 544}

\makeatother
\end{thebibliography}




\appendix

\section{Dependence of filament gas properties with halo contamination}
\label{app:halo_cont}

As supplementary material we include the result of the stacking procedure of Figs.~\ref{fig:Stack_all} and~\ref{fig:Stack_shocked}, but without excluding galaxy haloes. 
In general, we observe the same features and trends found in Section~\ref{sec:angular}, although including the disruption owing to haloes in the diffuse material of filaments making the profiles noisier and mainly affecting profiles in the inner cluster regions, as expected.

\begin{figure}
  \includegraphics[width=\columnwidth]{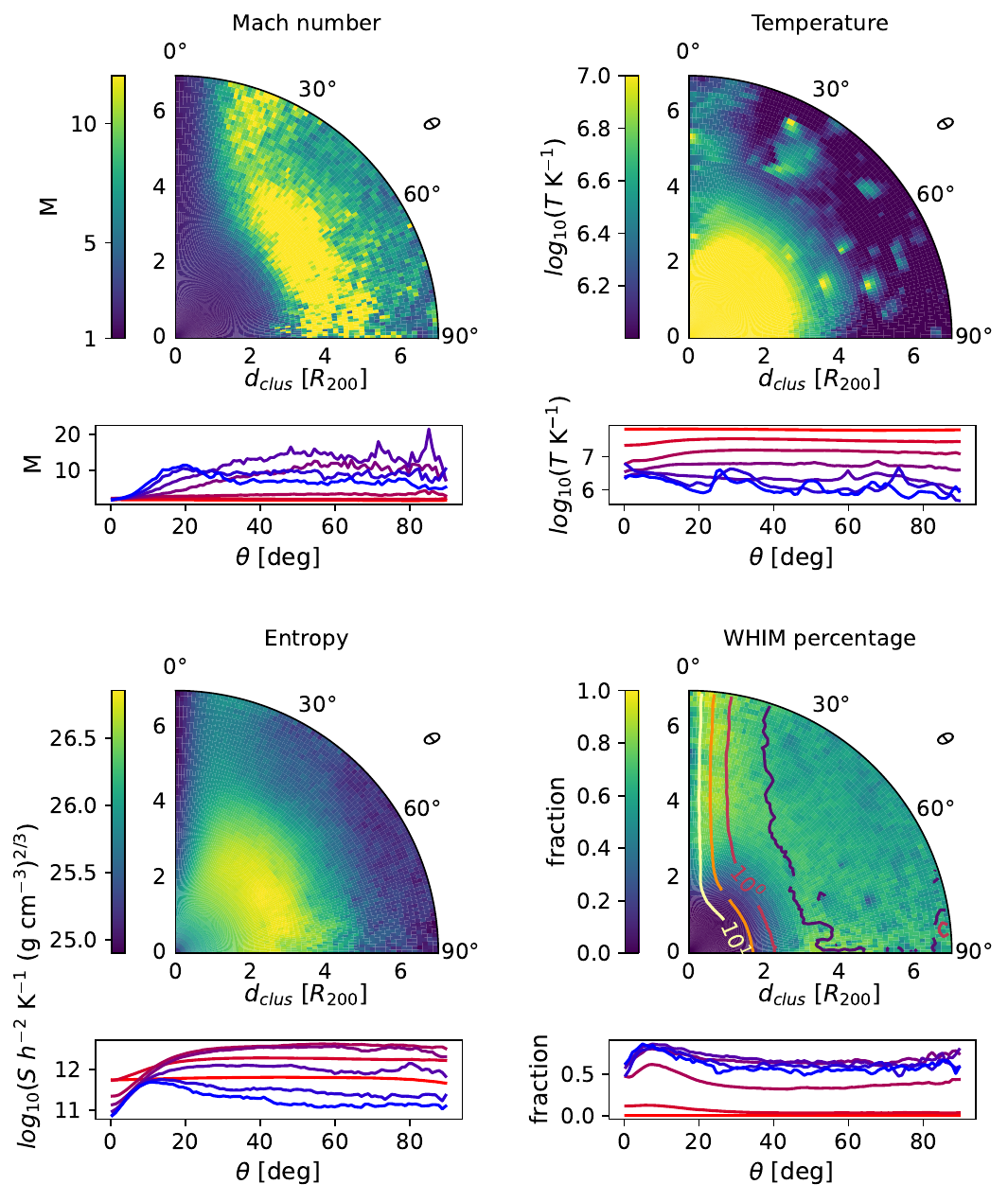}
  \caption{Same as Fig.~\ref{fig:Stack_all} but without masking out galaxy haloes. The 1D plots have been expanded with profiles inside the previously masked out within $2R_{200}$ including now two more layers corresponding to the ranges $0 \leq d_{\rm clus} < R_{200}$ and $R_{200} \leq d_{\rm clus} < 2 R_{200}$.}
  \label{fig:ApenStack0}
\end{figure}

\begin{figure}
  \includegraphics[width=\columnwidth]{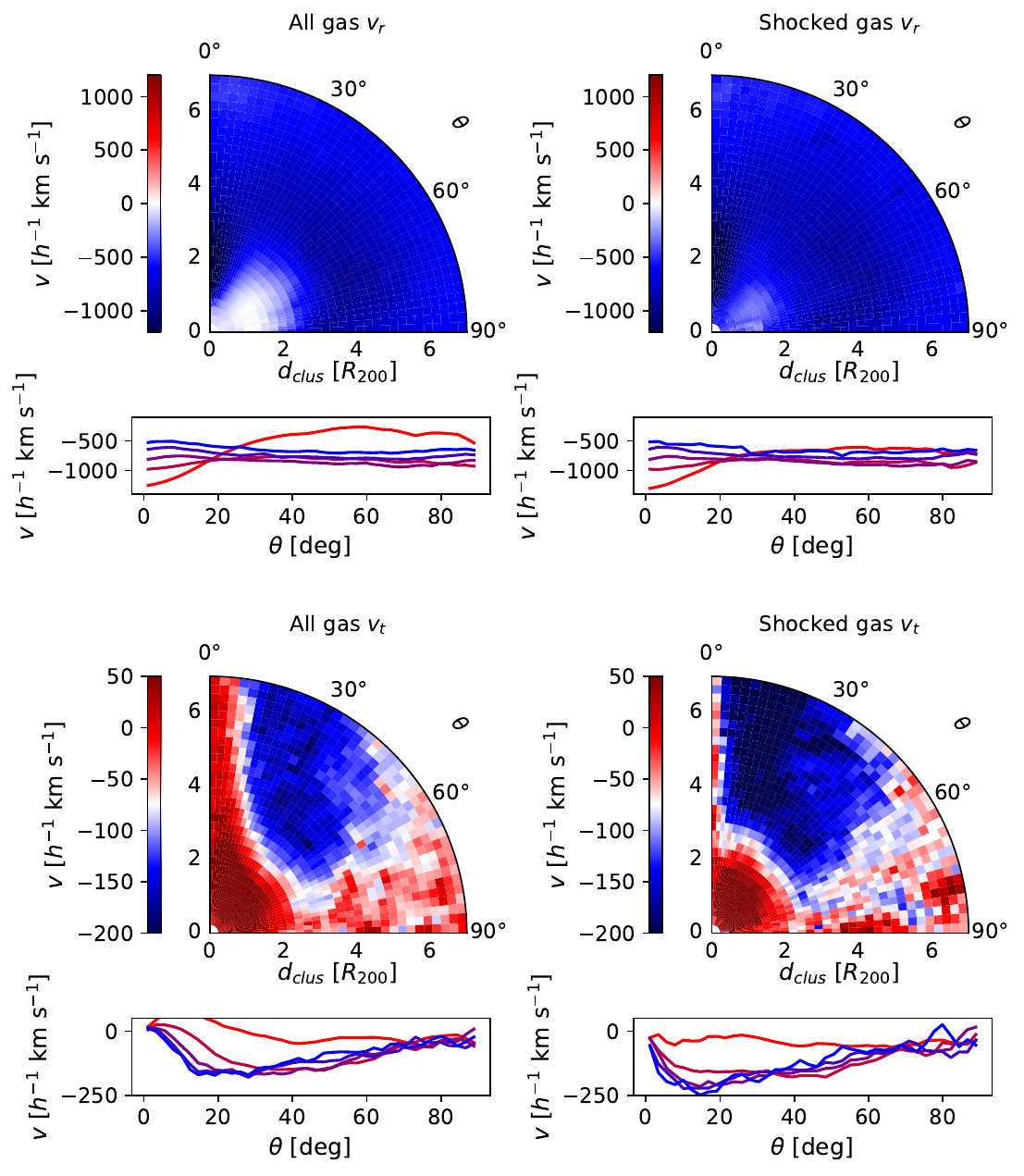}
  \caption{Same as Fig.~\ref{fig:Stack_shocked} but without masking out galaxy haloes. The 1D profiles include now two more layers corresponding to the ranges $0 \leq d_{\rm clus} < R_{200}$ and $R_{200} \leq d_{\rm clus} < 2 R_{200}$.}
  \label{fig:ApenStack1}
\end{figure}


\bsp	
\label{lastpage}
\end{document}